\documentclass[pre,aps,floats,superscriptaddress,floatfix,twocolumn]{revtex4}
\usepackage{amssymb,amsmath}
\usepackage{amsmath,amssymb}
\usepackage{graphicx}
\usepackage{psfrag}
\usepackage{color}
\usepackage{dcolumn}
\usepackage{bm}
\usepackage[normalem]{ulem}

\def\beq{\begin{equation}}
\def\eeq{\end{equation}}
\def\bea{\begin{eqnarray}}
\def\eea{\end{eqnarray}}
\begin{document}
  \title{Scaling or multiscaling: Varieties of universality in a driven nonlinear model}
\author{Sudip Mukherjee}\email{sudip.bat@gmail.com, sudip.mukherjee@saha.ac.in}
\affiliation{Barasat Government College,
10, KNC Road, Gupta Colony, Barasat, Kolkata 700124,
West Bengal, India}
\affiliation{Condensed Matter Physics Division, Saha Institute of
Nuclear Physics, Calcutta 700064, West Bengal, India}
\author{Abhik Basu}\email{abhik.123@gmail.com, abhik.basu@saha.ac.in}
\affiliation{Condensed Matter Physics Division, Saha Institute of
Nuclear Physics, Calcutta 700064, West Bengal, India}

\date{\today}
\begin{abstract}
Physical understanding of how the interplay between symmetries and nonlinear effects can control the scaling and multiscaling properties in a coupled driven system, such as magnetohydrodynamic turbulence or turbulent binary fluid mixtures, remains elusive till the date. To address this generic issue, we construct a conceptual nonlinear hydrodynamic model, parametrised jointly by the nonlinear coefficients, and the spatial scaling 
of the variances of the advecting stochastic velocity and 
the stochastic additive driving force, respectively. By using a perturbative one-loop dynamic 
renormalisation group method, 
we  calculate the 
 multiscaling exponents of the suitably defined equal-time structure functions of the dynamical variable. We 
show that depending upon the control parameters the model can display a variety of  universal scaling behaviours ranging from simple scaling to multiscaling.

 \end{abstract}

 \maketitle
 \section{Introduction}\label{intro}

In contrast to the conventional driven diffusive models~\cite{ddlg}, the universality of the velocity fluctuations in the nonequilibrium steady state (NESS) of forced three-dimensional ($3d$) homogeneous and isotropic fully developed fluid turbulence are characterised by 
the {\em multiscaling} of equal-time 
structure functions ${\mathcal S}_{n}^v(r)$ of the longitudinal component of 
the 
velocity increments $\Delta v(r)=\hat {\bf r}\cdot[{\bf v}({\bf x+r},t) - {\bf 
v}({\bf x},t)]$, where $\hat {\bf r}$ is the unit vector along $\bf r$, the separation vector between the two space points:
${\mathcal 
S}_{n}^v(r)$ is defined as
\begin{equation}
 {\mathcal S}_n^v (r)=\langle |\Delta v ({\bf r})|^n\rangle,\label{vel-struc}
\end{equation}
for {\em all} positive integer $n$;
the separation $r$ belongs to the inertial range: $L\gg r\gg \eta_d$, where $L$ and $\eta_d$, respectively, are the integral scale or the forcing scale ($\sim$ system size) and small scale dissipation scale, that depends on the fluid viscosity.
Dimensional arguments due to Kolmogorov (K41)~\cite{k41} predicted that ${\mathcal S}_n^v (r)$ in 
homogeneous and isotropic fully developed turbulence should be independent of 
both $L$ and $\eta_d$ in the inertial regime $L\gg r\gg \eta_d$ and display 
{\em universal scaling}, $ {\mathcal S}_n^v (r)\sim 
r^{\zeta_n^v},\,\zeta_n^v=n/3$ is positive, and has a linear dependence  on $n$, known as {\em 
simple scaling}.
Later on, detailed  experimental and numerical studied have clearly shown the existence of  
corrections nonlinear in $n$ to these scaling, making $\zeta_n^v$ depend nonlinearly upon $n$. This is the essence of {\em 
multiscaling}~\cite{frisch,rahulrev}. Closed analytical forms for these nonlinear functions $\zeta_n^v$ are still unknown. 
 In spite of the extensive experimental and numerical results, a 
self-consistent microscopic theory for multiscaling starting from the forced 
Navier-Stokes equation is
 still lacking. { Renormalisation group (RG) methods designed to extract universal scaling exponents  
in critical phenomena 
and critical dynamics~\cite{zinn} as well as in nonequilibrium driven diffusive systems~\cite{ddlg}, are so far less successful in 
calculating the universal multiscaling in fully developed fluid 
turbulence; see Ref.~\cite{rev1} for detailed expositions on RG approaches to fluid turbulence.}

Difficulties in the analytical progress  of fluid turbulence studies have inspired scientists to 
construct and study simpler models, amenable to analytical methods, that might display qualitatively similar universal scaling behaviour. One such well-known attempt is the studies of multiscaling in the passive scalar turbulence model. In this model,
 a scalar $\theta$ (e.g., a concentration field) is passively advected by a velocity field $\bf v$ without affecting the dynamics of the flow field itself, i.e., the fluid motion remains autonomous, independent of the 
passively advected $\theta$. The generic form of the equation of motion of $\theta$ reads
\begin{equation}
 \frac{\partial\theta}{\partial t} + {\bf v}\cdot {\boldsymbol\nabla}\theta = \nu_\theta\nabla^2 \theta + f_\theta,
\end{equation}
where, $\nu_\theta$ is a diffusivity and $f_\theta$ is an additive noise.
In the well-known {\em Kraichnan model} for passive 
scalar turbulence~\cite{obu,kraich}, $\bf v$ is chosen to be  zero-mean Gaussian-distributed with a variance that is spatially long-ranged but 
temporally $\delta$-correlated, instead of being 
 solutions of the forced Navier-Stokes equation. This effectively reduces 
the problem to a 
theory linear in $\theta$ for a given 
velocity, which in turn makes the problem analytically 
tractable. 
%This model has been extensively studied by a variety of analytical 
%means, ranging from field-theoretic perturbation theories~\cite{adjhem} to 
%zero-mode 
%analyses~\cite{kupi} and non-perturbative methods~\cite{pagani} among others, 
%which yield for the scaling of the equal-time, even order 
%structure concentration
%functions  
Extensive studies on the passive scalar turbulence~\cite{adjhem,kupi,pagani} show universal scaling behaviour by the appropriate structure functions of $\theta$:
\begin{equation}
 {\mathcal S}_{2n}(r)\equiv \langle [\theta ({\bf x+r})-\theta({\bf 
x})]^{2n}\rangle \sim r^{\zeta_{n}}\label{pass-struc}
\end{equation}
for positive integers $n$,
where  the (equal) time labels of $\theta$ are suppressed;  $r=|{\bf r}|$ in the 
inertial range. The scaling exponents 
$\zeta_{2n}$ are found
to be nonlinear functions of order $n$, similar to the multiscaling in fluid 
turbulence. Unsurprisingly, the odd order structure functions vanish identically, 
since
 the Kraichnan model depends linearly on $\theta$. This remains true even when the passive scalar is advected by velocity fields obtained from the isotropically Navier-Stokes equation, instead of a Gaussian-distributed velocity~\cite{sreeni1}.

In 
the Kraichnan model, the variance of the external (additive) stochastic force is nonvanishing  and concentrated only at the largest scales. The 
zero-mean, incompressible, Gaussian-distributed velocity field ${\bf v}({\bf r},t)$ has a variance given by~\cite{kupi,adjhem}
\begin{equation}
 \langle v_i ({\bf q},t)v_j({\bf -q},0)\rangle = \frac{\tilde D  P_{ij}({\bf 
q})\delta (t)}{(q^2 + 1/L^2)^{\tilde\epsilon}};\label{vcorrold}
\end{equation}
 see also 
Ref.~\cite{others}. { Here, }  
$\tilde D$ is a constant,
$\bf q$ is a Fourier wavevector and 
$P_{ij}=\delta_{ij} - q_iq_j/q^2$ is the transverse projection operator. In the lowest order perturbative expansion in $\tilde\epsilon$, the 
multiscaling exponents are given by
\begin{equation}
 \zeta_{n}=2n - \frac{n\tilde\epsilon (d+2n)}{d+2},\label{adjh}
\end{equation}
 { where $d$ is the space dimension}. 
%This has been studied numerically as well, see, e.g., Refs.~\cite{num1,num2}.
%The Kraichnan passive scalar model has been 
%subsequently extended to include various different effects, e.g, compressibility 
%of the fluid~\cite{adjhem1}, effects of a mean gradient~\cite{gauding}, effects 
%of shear flows~\cite{antonia} and random shear flows~\cite{shear1}.

A major difference between homogeneous and isotropic fluid turbulence and the passive scalar turbulence problem is the linearity of the latter problem in $\theta$. Since the fully developed turbulence is a state where the advective nonlinear term is {\em relevant} (in a RG sense), such that the Reynolds number approaches infinity, it is generally believed that nonlinearity plays a crucial role in the ensuing multiscaling. Due to the inherent linearity of the problem, studies on the passive scalar turbulence models cannot shed much light in this regard. In the absence of any general theoretical framework and due to the considerable theoretical and technical challenges to study and extract the scaling behaviour in fully developed turbulence in a systematic manner  starting from the forced Navier-Stokes equation, it is instructive to construct simple driven models, which may be amenable to analytical treatments.
With this  viewpoint, in the present work, we revisit the problem of scaling and 
multiscaling in a simple conceptual nonlinear model. We construct this model by generalising the well-known Kraichnan model for passively advected substance concentration~\cite{obu,kraich} for a conserved three-component spin vector field $\phi_\alpha,\,\alpha=1,2,3$, driven by a 
stochastic 
advecting velocity $\bf v$ and an additive force $f_i$. We introduce nonlinear effects, by which we mean nonlinear interactions between the components of $\phi_\alpha$, and their effects on the dynamics of $\phi_\alpha$. The nonlinear effects are modeled by certain {\em mode-coupling terms} of the form well-known in the dynamics of classical Heisenberg magnets~\cite{chaikin}. The dynamics
of $\phi_\alpha$ is controlled jointly by $y$ and $\overline y$, spatial 
scaling exponents of the variances, respectively, of $\bf v$ and 
$f_i$.  
By 
using Wilson 
momentum shell dynamic 
renormalisation group (DRG) and within a one-loop perturbation approximation, 
we elucidate scaling and multiscaling in these models. We show that the new mode coupling term breaks the symmetry under a constant shift of $\phi_\alpha({\bf x},t)$, present in the original passive scalar turbulence model. We argue that as a result the appropriate structure functions that are invariant under same transformations as the underlying equation of motion and can characterise the universal scaling or multiscaling are of the form
\begin{equation}
 {\cal \tilde S}_{2n}(r)\equiv \langle \left[\phi_\alpha({\bf x+r})\phi_\alpha({\bf x})\right]^n\rangle \sim  r^{\tilde \zeta_{n}}
\end{equation}
in the parameter regime where the coupling constant of the mode coupling term is {\em relevant} (in a RG sense) that in turn is controlled by $y$ and $\overline y$. By using a dynamic RG framework, we show that the relevant structure functions ${\cal \tilde S}_{2n}(r)$  display {\em only simple scaling} but {\em no} multiscaling. In the phase space region, where the mode coupling term is irrelevant, the model effectively becomes linear in the long wavelength limit and essentially reduces to the Kraichnan passive scalar model. This reveals sensitive dependence of multiscaling properties (or, the lack of it) on the nonlinear effects, a key qualitative outcome of the present study. 

We calculate the multiscaling exponents $\tilde \zeta_{2n}$ that depend 
linearly (in case of simple scaling) or nonlinearly (in case of multiscaling)
on 
$n$ and are parametrised by $y$ and $\overline y$. We also show that in the 
inertial range,
${\cal \tilde S}_{2n}(r)$ explicitly depends on $L$ that ultimately leads to 
multiscaling (or lack thereof).
We establish 
the crucial role played by {\em both} the advecting velocity and the additive 
noise in the dynamical 
equation for 
$\phi_\alpha$. We, in particular, show how the spatial scaling of the variance of 
the 
additive stochastic force affects $\zeta_{2n}$. Our calculational framework 
directly 
extends the standard DRG calculations for scaling in driven diffusive 
models~\cite{ddlg}.
The rest of this article is organised as follows: In Sec.~\ref{model}, for our purposes we 
introduce two similar models, Model I and Model II that differ in the variance of the Gaussian distributed $\bf v$. We define the structure functions, that we use to characterise the multiscaling, in Sec.~\ref{struct-func}. Next, in Sec.~\ref{lin-scal}, we calculate the exact scaling behaviour in the linear limit of the model. 
Then in Sec.~\ref{rg},
we perform the renormalisation group analysis for the relevant model parameters of Model I.  In Sec.~\ref{lin-ope}, we analyse the scaling in the linear model from the perspective of OPE.
Next, in Sec.~\ref{multi}  we calculate the scaling and multiscaling exponents in Model 
I. We demonstrate that the nonlinear mode coupling terms can make the multiscaling observed without these nonlinearities (i.e., in the usual passive scalar model) disappear; instead, simple scaling of a particular type is predicted to be displayed. In Sec.~\ref{summ} we summarise and conclude. We provide some 
technical details and briefly discuss Model II in Appendix for the interested reader.

 \section{Nonlinear hydrodynamic model}\label{model}

 Let $\phi_\alpha({\bf x},t)$ be a three-component ``spin-vector'' that is being advected by a velocity field $v_i({\bf x},t)$. Here, $\alpha=1,2,3$ and $i=1,..,d$, where $d$ is the dimensionality of the physical space. The spins span an order parameter space, distinct from the physical coordinate space; spin index $\alpha$ have no relation with the Cartesian component index $i$. For notational consistency, we reserve Greek letters to denote the spin components; Roman letters are used to denote the Cartesian components. The equation of motion for $\phi_\alpha({\bf x},t)$ reads
 \begin{equation}
  \frac{\partial\phi_\alpha({\bf x},t)}{\partial t} + g\epsilon_{\alpha\beta\gamma}\phi_\beta\nabla^2 \phi_\gamma + \lambda {\bf v}\cdot{\boldsymbol\nabla}\phi_\alpha = \nu\nabla^2\phi_\alpha + \overline\eta_\alpha.\label{phieq}
 \end{equation}
Here, $\epsilon_{\alpha\beta\gamma}$ is totally antisymmetric three-dimensional tensor, $g$ and $\lambda$ are two coupling constants and can be of either sign, and $\nu$ is a diffusion constant~\cite{nu-comm}. Evidently, through the nonlinear mode-coupling term $g\epsilon_{\alpha\beta\gamma}\phi_\beta\nabla^2 \phi_\gamma$ different components of $\phi_\alpha$ interact. Notice that the term $\epsilon_{\alpha\beta\gamma}\phi_\beta\nabla^2 \phi_\gamma= \epsilon_{\alpha\beta\gamma} {\boldsymbol\nabla}\cdot(\phi_\beta  {\boldsymbol\nabla} \phi_\gamma)$, since $ ({\boldsymbol\nabla}{\boldsymbol \phi})\times ({\boldsymbol\nabla}{\boldsymbol \phi})=0$ identically. Stochastic function $\overline\eta_\alpha$ is assumed to be Gaussian distributed with zero mean and a variance given by (in the Fourier space)
\begin{equation}
 \langle\overline\eta_\alpha({\bf q},\omega)\overline\eta_\beta(-{\bf q},-\omega)\rangle=2\overline D|q|^{-\overline y},
\end{equation}
where $\overline D>0$ fixes the amplitude; ${\bf q}$ and $\omega$ are Fourier wavevector and frequency, respectively. We are generally interested in $\overline{y}>0$, corresponding to a spatially long-ranged noise in (\ref{phieq}).
When $\lambda=0$, Eq.~(\ref{phieq}) reduces to that of the well-known Model E equation for the classical Heisenberg model in its paramagnetic phase (i.e., above the transition temperature)~\cite{halpin,u-term}. The $\lambda {\bf v}\cdot {\boldsymbol\nabla}\phi_\alpha$ describes advection of the spins by the velocity $v_i$. 

For a realistic 3D system, the incompressible velocity $v_i$ solves  the generalised Navier-Stokes equation that includes the stresses coming from the spin fluctuations. Such velocity field itself, when strongly forced externally at large spatial scales, displays nontrivial multiscaling or intermittency. There are driven coupled system, where such an intermittent flow field advects another field like the magnetic field in 3D MHD turbulence, or the concentration field in a turbulent binary mixture. Any theoretical analysis of the scaling or multiscaling of the second field is greatly complicated by the multiscaling of the advecting velocity field.  In the spirit of the Kraichnan passive scalar model and in order to keep the ensuing calculation simple and analytically tractable, instead of solving the velocity field from the Navier-Stokes equation, it is assumed as an ``input'' as a zero-mean 
Gaussian 
distributed field with a given variance that is spatially long-ranged. 
 Thus the velocity field in this description  is {\em autonomous}, and has a variance
$D^v_{ij}({\bf q},\omega)$  in the Fourier space:
\begin{equation}
 \langle v_i({\bf q},\omega) v_j({\bf -q},-\omega)\rangle = D_{ij}^v({\bf 
q},\omega).\label{variv-gen}
\end{equation}
Equivalently, in the time domain
\begin{equation}
\langle v_i ({\bf q},t)v_j ({\bf -q},0)\rangle=D_{ij}^v({\bf q},t).
\end{equation}
In 
particular, we assume
\begin{equation}
  D_{ij}^v({\bf q},t) = A P_{ij}({\bf q}) q^{-y} 
 {\rm exp}(-\tilde\Gamma (q) t),\label{vcorr}
\end{equation}
where, $A>0$ is a constant that sets the amplitude of the variance 
$D_{ij}^v$, 
and the exponent $y>0$; $P_{ij}({\bf q})\equiv \delta_{ij}-q_iq_j/q^2$ is the transverse projection operator. Parameter $\tilde\Gamma (q) >0$ controls the temporal decay of 
the time-dependent velocity correlator { and parametrises (\ref{vcorr})}. 
We consider two specific choices for $\tilde\Gamma (q)$~\cite{tirtha}:

(i) Model I: $\tilde\Gamma(q)=\Gamma$, a constant (independent of $q$) together with $\Gamma \rightarrow 
\infty$, along with the condition $A$ scales with $\Gamma$, i.e., $A/\Gamma = D_1>0$ a const.: 
relaxation of the velocity modes are 
independent of wavevector $\bf q$. In that limit, (\ref{vcorr}) reduces to 
being temporally $\delta$-correlated:
\begin{equation}
 D_{ij}^v({\bf q},t) = D_1P_{ij}({\bf q}) q^{-y}\delta 
(t),\label{model1v}
 \end{equation}
  Such a flow can exist when there is  a (large) friction (e.g., through a porous medium)
and a large external stirring force applied on the fluid, so that the friction and the force may balance and produce a nonequilibrium steady state. This in turn 
produces a flow field correlated as in (\ref{model1v}). The flow is 
self-similar, which is evident from (\ref{model1v}); it is however, { not {\em intermittent}, since it is Gaussian-distributed, in contrast to turbulent velocity 
fields that satisfy the forced Navier-Stokes equation~\cite{frisch}}.
% This has been used in the literature already~\cite{adjhem,others}. { 
%Notice that Eq.~(\ref{thetaeq}) in conjunction  with variances 
%(\ref{noisevari}) and (\ref{model1v}) is invariant under the {\em Galilean 
%transformation}: ${\bf x^\prime}={\bf x}+ {\bf c} t,\, t^\prime = t,\, 
%\partial/\partial t^\prime = \partial/\partial t - \lambda {\bf c}\cdot 
%{\boldsymbol \nabla}$. Here, $\bf c$ is the Galilean boost.}
%We 
%elucidate below how the universal scaling behaviour is parametrised by $\overline y, 
%y$. In 
%particular, we show below that the choice 
%$\overline y=d$ together with (\ref{model1v}) above reproduce the results on 
%the multiscaling of ${\mathcal S}_{2n}(r)$ from the well-known Kraichnan model 
%for passive scalar~\cite{obu,kraich}.
%\begin{equation}
% \langle v_i({\bf q},\omega) v_j({\bf -q},-\omega)\rangle \propto  P_{ij}({\bf 
%q})|q|^{-y}.\label{variv}
%\end{equation}
%Thus the velocity field in this description has no dynamics and is {\em autonomous}, i.e., independent of anything else. 

(ii) Model II:  $\tilde\Gamma=\eta q^2$.  This is equivalent to $\bf v$ satisfying 
the linearised Navier-Stokes equation:
\begin{equation}
 \frac{\partial v_i}{\partial t}=\eta \nabla^2 v_i-{\nabla_i P}  
+\tilde f_i,\label{ns1}
\end{equation}
with $\langle v_i \rangle=0$; $P$ is the pressure and $\eta$ the kinematic 
viscosity. Also, ${\bf v}$ is 
assumed to be incompressible, i.e., ${\boldsymbol \nabla}\cdot{\bf v}=0$. We use this  to eliminate $P$ from (\ref{ns1}), and get
\begin{equation}
 \frac{\partial v_i}{\partial t}=\eta_v \nabla^2 v_i +P_{ij}g_j.\label{ns2}
\end{equation}
 Here, $\eta_v$ is a fluid viscosity. Function $\tilde f_i$ is a zero-mean Gaussian 
distributed stochastic force with a variance
\begin{equation}
\langle \tilde f_i ({\bf q},t) \tilde f_j ({-\bf q},0)\rangle = \tilde D_1|q|^{-y}\delta 
(t)\delta_{ij}, 
\end{equation}
where subscripts $i,j$ refer to Cartesian coordinates; $\tilde D_1>0$. { This 
yields
\begin{equation}
 D^v_{ij}({\bf q},\omega)=\frac{2\tilde D_1 |q|^{-y} P_{ij}({\bf q})}{\omega^2 
+ \eta^2 
q^4}.
\end{equation}
We are again interested in the limit of large $\tilde\Gamma$, which in this case implies a large $\eta_v$, with $\tilde D_1/\eta_v=D_\eta$ is finite. This is the so-called Stokesian limit of the Navier-Stokes equation.

\section{Structure functions}\label{struct-func}

In both the problems of fully developed turbulence and passive scalar turbulence, the forms (\ref{vel-struc}) and (\ref{pass-struc}), respectively, of the structure functions are guided by the symmetry of the underlying equations of motion themselves. In the same spirit, we note that the only symmetry of $\phi_\alpha$ that keeps Eqs.~(\ref{phieq}) and (\ref{vcorr}) invariant is the rotation in the order parameter space. This prompted us to define the structure function
\begin{equation}
  {\cal \tilde S}_{2n}(r)\equiv\langle \left[\phi_\alpha({\bf x+r})\phi_\alpha({\bf x})\right]^n\rangle \sim r^{\overline\zeta_n}.\label{struct1}
\end{equation}
Here, $n$ is any positive integer.  Rotational invariance ensures that the any odd order analogues of the structure functions (\ref{struct1}) should vanish. Thus only even order structure functions are considered. On the other hand, in a subspace of the full parameter space where the coupling constant $g=0$, i.e., the mode coupling term (\ref{phieq}) vanishes, the reduced system has a higher symmetry: In addition to the rotational invariance, it is also invariant under a constant shift $\phi_\alpha \rightarrow \phi_\alpha +\,const.$. This allowed us to define yet another class of structure functions ${\cal S}_{2n}(r)$ as
\begin{equation}
 {\cal S}_{2n}(r)\equiv \langle [\phi_\alpha({\bf x+r})-\phi_\alpha({\bf x})]^{2n}\rangle \sim r^{\zeta_n}.
\end{equation}
These are same as those used in the standard passive scalar turbulence problem as given in (\ref{pass-struc}) above~\cite{tirtha}.
Here too, the odd order analogues of ${\cal S}_{2n}(r)$ vanish. 

\section{Scaling in the linear theory}\label{lin-scal}

Before embarking on investigating the scaling properties of the full nonlinear model, it is instructive to consider the same for the linearised version of the model, i.e., with $g=0=\lambda$. Thus the dynamics of $\phi_\alpha$ is decoupled from $v_i$ and just follows the linear equation
\begin{equation}
 \frac{\partial \phi_\alpha}{\partial t} = \nu\nabla^2 \phi_\alpha + \overline\eta_\alpha.\label{lin-eq}
\end{equation}
In this section we closely follow the discussions in Ref.~\cite{tirtha}. 
By using (\ref{lin-eq}), we obtain in the Fourier space
\begin{equation}
 \langle\phi_\alpha ({\bf k},\omega)\phi_\beta(-{\bf k},-\omega)=\frac{2\overline Dk^{-\overline y}\delta_{\alpha\beta}}{\omega^2 + \nu^2 k^4}.
\end{equation}
Linearity of the dynamics of $\phi_\alpha$ ensures that all order correlation and structure functions can be calculated {\em exactly}.

Following Ref.~\cite{tirtha}, we write
\begin{eqnarray}
 &&{\cal \tilde S}_2(r)\equiv \langle \phi_\alpha({\bf x+ r},t)\phi_\alpha({\bf x},t)\rangle\nonumber \\ &=&\int_{2\pi/L}^\Lambda \frac{d^dq}{(2\pi)^d}\frac{\overline D\delta_{\alpha\alpha}}{q^{2+\overline y}}\exp(i{\bf q\cdot r}).\label{cal-tilde-s2}
\end{eqnarray}
For $\overline y+2<d$, (\ref{cal-tilde-s2}) is insensitive to the lower limit, which may be brought to zero (i.e., $L\rightarrow\infty$) without any problem. This gives
\begin{equation}
 {\cal \tilde S}_2\sim r^{\overline y+2-d}.
\end{equation}
The scaling of the higher order structure functions ${\cal \tilde S}_{2n}(r)$ can be readily found by noting that due to the linearity of the dynamics $\phi_\alpha$ is Gaussian distributed. This gives
\begin{equation}
 {\cal \tilde S}_{2n}(r)\sim r^{n(\overline y+2-d)}.\label{exact11}
\end{equation}
This includes the case $\overline y=-2$, i.e., a conserved noise in (\ref{phieq}). Equation~(\ref{exact11}) shows that ${\cal \tilde S}_{2n}(r)$ is a decaying function of $r$.
On the other hand, for $\overline y+2>d$, (\ref{cal-tilde-s2}) is dominated by the lower limit, giving
\begin{equation}
 {\cal\tilde S}_2(r)= \int_{2\pi/L}^{\Lambda} \frac{d^dq}{(2\pi)^d}\frac{D_1\delta_{\alpha\alpha}}{q^{2+\overline{y}}}\sim L^{\overline{y}+2-d}\label{ans1}
\end{equation}
in the asymptotic limit $L\gg r$. Thus, ${\cal \tilde S}_2(r)$ is {\em independent} of $r$, a growing function of $L$ and in fact diverges as $L\rightarrow \infty$ with $L\gg r$. Similarly, in the linearised theory using the logic outlined above, and by employing Gaussianisation (exact for a linear dynamical system) we find
\begin{equation}
 {\cal \tilde S}_{2n}(r)\sim L^{n(\overline{y}+2-d)}\label{exact12}
\end{equation}
independent of $r$ and diverges with $L$, since $\overline y+2>d$.

In the same way,
\begin{eqnarray}
 &&{\cal S}_2(r)\equiv 2\langle \phi_\alpha({\bf x},t)\phi_\alpha({\bf x},t)\rangle - 2 \langle \phi_\alpha ({\bf x},t)\phi_\alpha(0,t)\rangle \nonumber \\&=& 2\int_{2\pi/L}^\Lambda \frac{d^dq}{(2\pi)^d}\frac{\overline D\delta_{\alpha\alpha}}{q^{2+\overline y}}\left[1-\exp(i{\bf q\cdot r})\right].
\end{eqnarray}
This gives, as in Ref.~\cite{tirtha}, for $\overline y<d$,
\begin{equation}
  {\cal S}_2(r) \sim r^{2-d+\overline{y}},\,{\cal S}_{2n}(r) \sim r^{n(2-d+\overline{y})},
\end{equation}
where as for $\overline y>d$
\begin{equation}
 {\cal S}_2(r) \sim r^2L^{\overline y-d},\,{\cal S}_{2n}(r) \sim r^{2n}L^{n(\overline y-d)}.\label{lin-1}
\end{equation}
Unsurprisingly, in the linear model ${\cal S}_{2n}(r)$ shows only simple scaling~\cite{tirtha}.

We will see below that complex scaling behaviours by ${\cal \tilde S}_{2n}(r)$ and ${\cal S}_{2n}(r)$ emerge, when the nonlinear effects are taken into account. We address this systematically in the next Section.

\section{Nonlinear effects in Model I}\label{nonlin}

When the nonlinear terms are considered, $\phi_\alpha({\bf x},t)$ can no longer be solved exactly. This necessitates perturbative approaches. However, na\"ive perturbative expansions produce diverging corrections. These difficulties can be systematically handled within a renormalisation group calculation framework. 

The dynamic RG procedure is well-documented in the literature~\cite{drg}. We give here a brief outline of the method for the convenience of the readers. This starts by writing down the dynamic generating functional
\begin{equation}
 {\cal Z}=\int {\cal D} \phi_\alpha{\cal D}\hat\phi_\alpha{\cal D}v_i \exp [-S_{act}],
\end{equation}
where the action functional $S_{act}$ is given by
\begin{widetext}
\begin{eqnarray}
 { S}_{act} &=& -\int d^d q d\omega [\overline D q^{-\overline y} \hat \phi_\alpha ({\bf q},\omega) \hat 
\phi_\alpha ({\bf -q},-\omega)+ 
\hat\phi_\alpha ({\bf -q},-\omega) [-i\omega \phi_\alpha ({\bf q},\omega) + 
i\lambda q_m \sum_{{\bf q}_1,\Omega} v_m({\bf q}_1,\Omega) \phi_\alpha ({\bf 
q-q}_1,\omega -\Omega)\nonumber \\
 &+&\nu q^2 \phi_\alpha({\bf q},\omega)+{\bf q\cdot q}_1\phi_\beta({\bf q-q}_1,\omega-\Omega)\phi_\gamma({\bf q}_1,\Omega)]-v_i ({\bf q},\omega)
[D_{ij}^v(q,\omega)]^{-1}v_j({\bf -q},-\omega)/2].\label{action1}
\end{eqnarray}

\end{widetext}
Here, $\hat\phi_\alpha({\bf x},t)$ is the usual response field~\cite{janssen}.

The momentum shell dynamic RG consists of tracing over the short wavelength Fourier modes of $\phi_\alpha({\bf x},t),\,\hat\phi_\alpha({\bf x},t)$ and ${\bf v}({\bf x},t)$, followed by rescaling of lengths and time. We follow the standard procedure of initially restricting the wavevectors to lie in a $d$-dimensional Brillouin zone: $q\leq \Lambda$, where $\Lambda$ is an ultra-violet cutoff, which should be of the order of the inverse lattice spacing $a_0$. However, its precise value has no effect on the theory that we develop. The fields ${
\phi_\alpha}({\bf x},t),\,\hat\phi_\alpha({\bf x},t)$ and ${\bf v}({\bf x},t)$ 
are then separated into high and low wave vector parts:
$\phi_\alpha({\bf x},t)=\phi_\alpha^>({\bf x},t)+\phi_\alpha^<({\bf x},t),\,\hat\phi_\alpha({\bf x},t)=\hat\phi_\alpha^>({\bf x},t)+\hat\phi_\alpha^<({\bf x},t)$ and ${\bf v}({\bf x},t)={\bf v}^>({\bf x},t) + {\bf v}^<({\bf x},t)$,
where $\phi_\alpha^>({\bf x},t),\,\hat\phi_\alpha^>({\bf x},t)$ and ${\bf v}^>({\bf x},t)$ all have support in the large wave vector  (short wavelength) 
range $\Lambda/b<| q|<\Lambda$, while $\phi_\alpha^<({\bf x},t),\,\hat\phi_\alpha^<({\bf x},t)$ and ${\bf v}^<({\bf x},t)$ have support in the small 
wave vector (long wavelength) range $|{ q}|<\Lambda/b,\,b>1$. In the next step, we integrate over $\phi_\alpha^>({\bf x},t),\,\hat\phi_\alpha^>({\bf x},t)$ and ${\bf v}^>({\bf x},t)$, which can be done perturbatively in the anhamornic couplings in (\ref{action1}). The perturbation theory that is constructed here is represented by the Feynman diagrams, with the order of perturbation theory being given by the number of loops in the Feynman diagrams that we consider here. After this perturbative step, we are to rescale lengths with ${\bf x}=b{\bf x'},\,b>1$, so as to restore the UV cutoff to $\Lambda$ again, and time with $t=t'b^z$, where $z$ is the dynamic exponent. This is then followed by rescaling of $\phi_\alpha^< ({\bf x},t),\,\hat\phi_\alpha^< ({\bf x},t)$ and ${\bf v}^< ({\bf x},t)$. In particular, we rescale $\phi_\alpha^<({\bf x},t)$ by $b^\chi$, where the scaling exponent $\chi$ is used to produce a fixed point, and charactrises the spatial scaling of the auto-correlations of $\phi_\alpha({\bf x},t)$. 

%\section{Multiscaling in Model I}\label{mult-mod}

\subsection{Renormalisation of the model parameters}\label{rg}

The RG procedure allows us to obtain the fluctuation-corrected model parameters.  First we consider $\overline y = -2$. One can formally  construct a one-loop diagram for $\overline D$ that comes from the mode-coupling term. However, these one-loop fluctuation corrections to $\nu$ and $\overline D$ that come from the mode coupling nonlinear term are {\em not} infra-red divergent and hence are not relevant. Only the advective nonlinearity contributes to the relevant corrections to $\nu$ and $\overline D$~\cite{tirtha}. Unsurprisingly, our results for $\overline y =-2$ are identical to those in Ref.~\cite{tirtha}. We revisit those results here for the sake of completeness. The relevant one-loop Feynman graphs for $\nu$ are shown in Appendix~\ref{feyn-nu}, some of which come from the mode-coupling nonlinearity.
\begin{eqnarray}
 \nu^<&=& \nu\left[1+\frac{\lambda^2 D_1}{\nu}\left(1-\frac{1}{d}\right)k_d\frac{\Lambda^{d-y}}{d-y} \left(1-\frac{1}{b^{d-y}}\right)\right],\\
 \overline D^<&=&\overline D\left[1+\frac{\lambda^2 D_1}{\nu}\left(1-\frac{1}{d}\right)k_d\frac{\Lambda^{d-y}}{d-y} \left(1-\frac{1}{b^{d-y}}\right)\right],
\end{eqnarray}
where $k_d=S_d/(2\pi)^d$ with $S_d$ being the hypersurface area of a $d$-dimensional sphere of unit radius. Writing $b=\exp(\delta l)\approx 1+\delta l$, we obtain the flow equations
\begin{eqnarray}
 \frac{d\nu}{dl}&=& \nu\left[z-2+\frac{\lambda^2 D}{\nu}\left(1-\frac{1}{d}\right)k_d\Lambda^{d-y}\right],\\
 \frac{d\overline D}{dl}&=&\overline D\left[z-2\chi-d-2+\frac{\lambda^2 D}{\nu}\left(1-\frac{1}{d}\right)k_d\Lambda^{d-y}\right].\nonumber \\
\end{eqnarray}
Solution of the above flow equations yield $z=2$  and $\chi=-d/2$~\cite{tirtha}. Here, $\chi$ is the scaling dimension of the field $\phi_\alpha({\bf x},t)$.

We now focus on the case with $\overline y >-2$. With this, notice that there are no relevant (in a RG sense) corrections to $\overline D$.  This is because the corrections to $\overline D$ are all ${\cal O}(q^2)$, hence cannot generate a singular (in the IR limit) contribution that may correct $\overline D q^{-\overline y}$. This gives, as in Ref.~\cite{tirtha},
\begin{eqnarray}
 \frac{d\overline D}{dl}&=&\overline{D}\left[z-2\chi-d+\overline y\right],\label{flowD}
\end{eqnarray}
for all values of $\overline y$. Thus
\begin{equation}
 \chi=(\overline y + z-d)/2 \label{exact-expo}
\end{equation}
is an exact exponent identity. This in the linear theory reduces to $\chi=(\overline y + 2-d)/2$ with $z=2$.

There are relevant one-loop fluctuation corrections to $\nu$, however. The fluctuation-corrected diffusivity $\nu^<$ is given by 
\begin{eqnarray}
 \nu^<&=&\nu [1+\frac{\lambda^2D_1}{\nu d} (d-1)k_d \int dq q^{ -y + d-1}   \nonumber \\&+& \frac{g^2 \overline D \overline y}{2\nu^3 d}k_d \int dq q^{-\overline y + d-3}]\label{nu-correct}
\end{eqnarray}
see Appendix~\ref{feyn-nu} for detailed calculations.
Equation~(\ref{nu-correct}) allows us to define two effective coupling constants: $g_1= k_d \frac{g^2 \overline D \overline y}{\nu^3}\Lambda^{d-\overline y-2}$, $g_2=k_d\frac{\lambda^2D_1}{\nu }\Lambda^{d-y}$.% and $g_3=k_d \frac{\alpha D_1}{\nu^2 }$. 
 There are no fluctuation corrections to $g,\,\lambda,\,\overline D,\,D_1$. That $g$ does not renormalise is a consequence of the rotational invariance of the order parameter space.  There are actually one-loop graphical corrections to $\lambda$; these however produce subleading corrections (see Ref.~\cite{tirtha}).

Rescaling of the space, time and fields allow us to convert (\ref{nu-correct}) into a differential RG flow equation. We get
\begin{equation}
 \frac{d\nu}{dl}=\nu\left[z-2+\frac{g_1}{2d}+g_2 \frac{d-1}{d}\right].\label{flow-nu1}\\
\end{equation}

The RG flow equations for the coupling constants $g_1$ and $g_2$ are as follows:
\begin{eqnarray}
\frac{dg_1}{dl}&=&g_1\left[\overline y -d +2 - \frac{3g_1}{2d}-3g_2 \frac{d-1}{d}\right], \label{flow-g1}\\
 \frac{dg_2}{dl}&=&g_2\left[y-d -\frac{g_1}{2d}- g_2\frac{d-1}{d}\right].\label{flow-g2}
\end{eqnarray}
It is convenient to define two ``small parameters'' $\delta \equiv \overline y -d+2$ and $\epsilon \equiv y-d$. Clearly, $\delta $ and $\epsilon$, respectively, are the scaling dimensions of $g_1$ and $g_2$. In fact, the one-loop perturbation theory that we set up below is effectively an expansion in $\epsilon$ and $\delta$ up to the linear order in these two parameters.

The RG fixed points $(g^*_1,\,g^*_2)$ are the solutions of $dg_1/dl=0=dg_2/dl$. Flow Eqs.~(\ref{flow-g1}) and (\ref{flow-g2}) four sets of fixed points (FP):
\\\\
(i) FP1: $(g^*_1,\,g^*_2)=(0,\,0)$: this is linearly stable for $\overline y< d-2$, i.e., $\delta<0$ and  $y<d$, i.e., $ \epsilon<0$. At this fixed point, unsurprisingly, the dynamic exponent $z=2$, same as that in the linear theory.\\\\
(ii)  FP2: $(g^*_1,\,g^*_2)=(\frac{2d}{3}\delta,\,0)$: this is linearly stable for $\delta/3>\epsilon$. At this FP, 
\begin{equation}
z=2- (\overline y -d + 2)/3=2-\frac{\delta}{3}.
\end{equation}
\\
(iii)  FP3: $(g^*_1,\,g^*_2)=(0,\,\frac{d}{d-1}\epsilon)$: this is linearly stable for $\epsilon>\delta/3$. At this fixed point, $z=2- (y-d)=2-\epsilon$. Since $y>d$, i.e., $\epsilon>0$, at this FP, we have $z<2$. \\\\
(iv) Nontrivial FP: If $y-d=(\overline y -d +2)/3> 0$, { i.e., $\epsilon=\delta/3$} then
\begin{equation}
 \frac{g^*_1}{2}+g^*_2(d-1)=d(y-d)=(\overline y-d+2)\frac{d}{3}.
\end{equation}
  However, $g^*_2$ and $g^*_1$ cannot however be separately evaluated~\cite{foot2}. By using (\ref{flow-nu1}), we conclude that 
\begin{equation}
 z=2 - \frac{g^*_1}{d}-g^*_2\frac{d-1}{d}= 2-(y-d)<2.
\end{equation}
{
We find that 
\begin{equation}
 y-d=\left(\overline{y}-d+2\right)/3,\;\;{\rm i.e.}, \epsilon=\frac{\delta}{3}
\end{equation}
is the separatrix between the two fixed points in the $g_1-g_2$ plane.}
 We note that the systems with parameters satisfying $y-d> (\overline y -d +2)/3>0$ flow to the linearly stable fixed point 
\begin{equation}
 g^*_1=0,\,g^*_2=\frac{d}{d-1}(y-d).
\end{equation}
 This gives 
\begin{equation}
z=2-(y-d)=2-\epsilon<2.
\end{equation}
 However, all systems with  $2(\overline y -d +2)/3>y-d>0$ flow to the linearly stable fixed point
\begin{equation}
 g^*_1=\frac{2d}{3}(\overline y -d +2),\,g^*_2=0,\,.
\end{equation}
This gives $z=2-(\overline y -d +2)/3=2-\delta/3<2$}. 

\begin{figure}[htb]
\includegraphics[width=7cm]{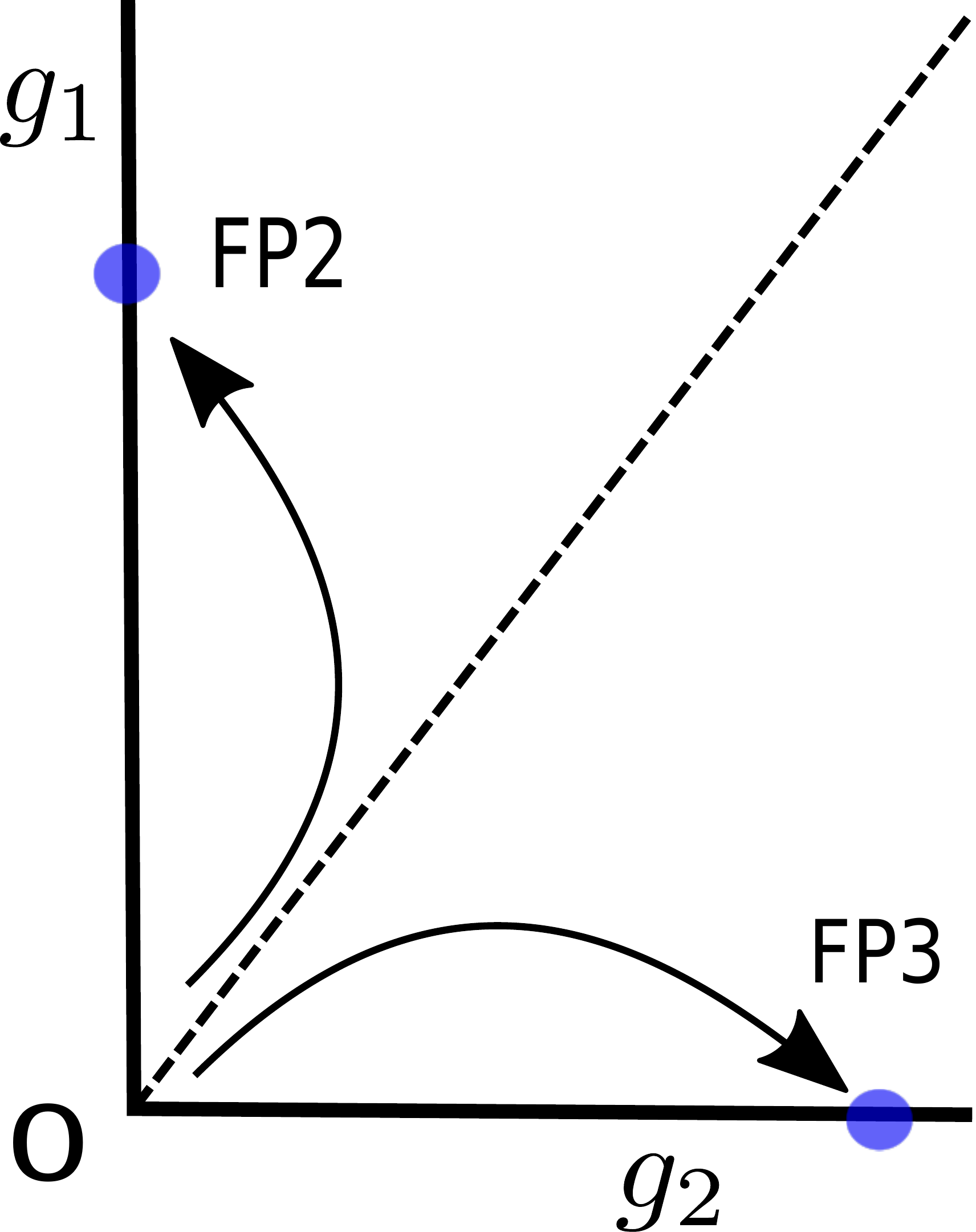}
\caption{RG flow diagram in the $g_2-g_1$ plane with $\overline y>-2$ and $y>0$.  FP2 and FP3 are the RG fixed points obtained above. The inclined broken line is the separatrix $y-d = (\overline y -d +2)/3$ for given $y,\overline y, d$.}\label{flow}\label{flow-diag}
 \end{figure}

See Fig.~\ref{flow-diag} for a schematic RG flow diagram in the $g_1-g_2$ plane.

\begin{figure}[htb]
\includegraphics[width=8cm]{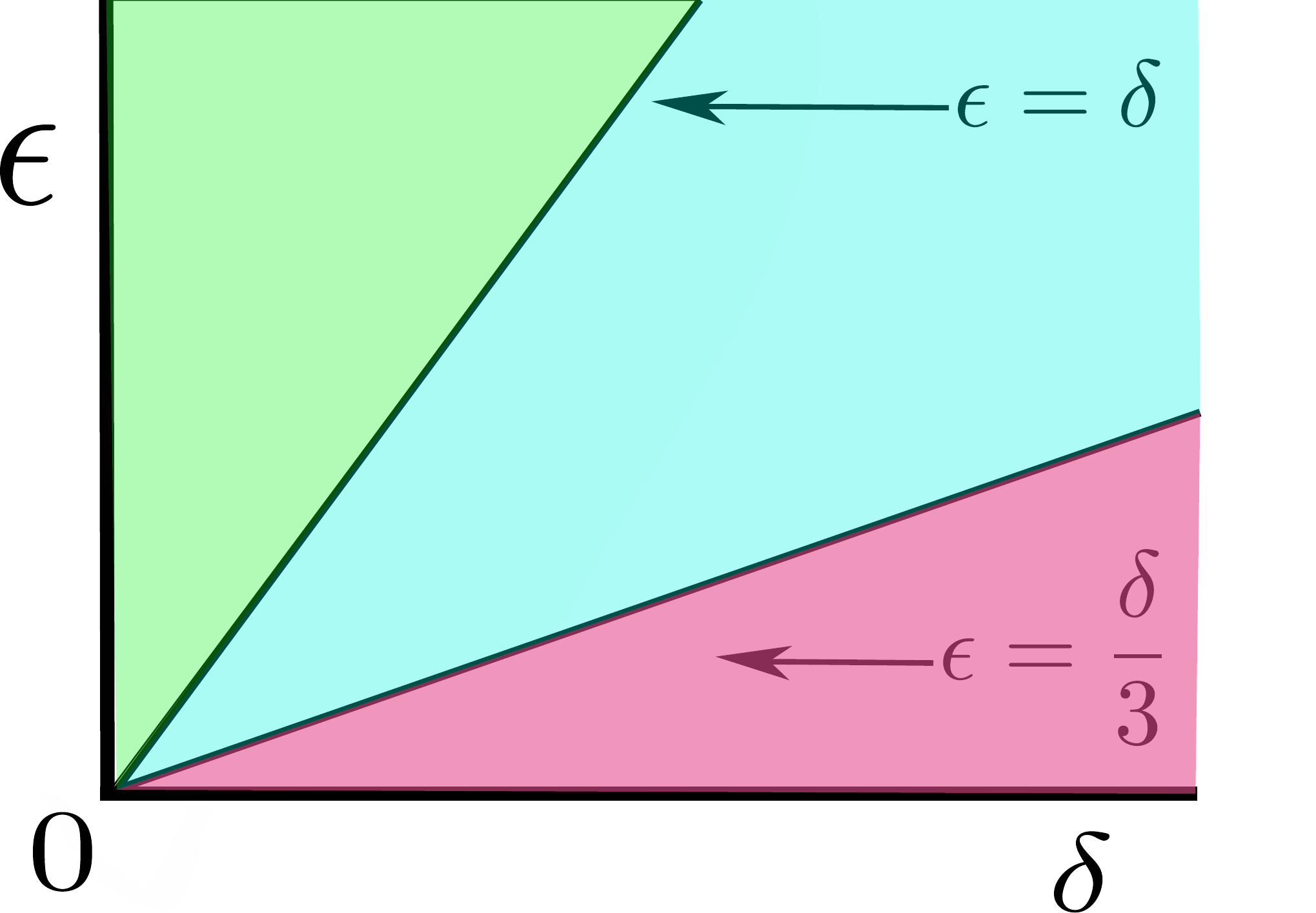}
\caption{Phase diagram of Model I in the $\delta-\epsilon$ plane. The mode-coupling term is relevant but the advective nonlinearity is irrelevant (in a RG sense) with $\chi>0$ in the renormalised theory in the lower purple triangle. In the intermediate light blue triangle, the mode coupling term is irrelevant, but the advective nonlinearity is now relevant with $\chi>0$ in the renormalised theory. In the light green upper triangle, $\chi<0$ (see text).  }\label{phase-diag}
 \end{figure}

We now obtain the spatial scaling exponent $\chi$. We note that the condition of the  nonrenormalisation of $\overline D$ (or equivalently the flow equation (\ref{flowD})) gives
\begin{equation}
\chi=(z+\overline y-d)/2,
\end{equation}
an exact relation that holds at the fixed points discussed above.
Since the value  of $z$ depends upon the stable fixed point, so does the value of $\chi$. For instance, when $g^*_1=0,\,g^*_2>0$, we get by using $z=2-(y-d)$
\begin{equation}
 \chi=(2-y+\overline y)/2.
\end{equation}
Here, $\chi$ can be made positive by tuning $y$ and $\overline y$. Assume $y-d\equiv \epsilon >0$ and $\overline y -d + 2\equiv \delta/2 >0$, with $\epsilon > \delta/3$. The last condition ensures that $g^*_2>0$ and $g^*_1=0$ at the FP. This gives
\begin{equation}
 \chi = \frac{1}{2}(2-y+\overline y)=\frac{1}{2}(\delta-\epsilon)>0.
\end{equation}
Thus, $\delta/3 <\epsilon <\delta$ is the necessary condition for $\chi>0$. On the other hand, when 
$g^*_2=0,\,g^*_1>0$, together with $z=(\overline y -d+2)/3$, we get 
\begin{equation}
 \chi = \frac{1}{2}(z+\overline y -d)=\frac{1}{2} (2-\frac{\delta}{3}+\delta-2)=\frac{2\delta}{3} >0
\end{equation}
along with $\delta/3 >\epsilon$. We find that the fluctuation corrections to the values of the scaling exponents $z$ and $\chi$ in the linear theory are linear in $\epsilon$ or $\delta$. We further note that at the stable fixed point FP2, $g_1^*\sim {\cal O}(\delta),\,g_2^*=0$, whereas at FP3, $g_1^*=0,\,g_2^*\sim{\cal O}(\epsilon)$, establishing the fact that the one-loop perturbation theory here is in effect an an expansion in $\epsilon$ and $\delta$. 
See Fig.~\ref{phase-diag} for a schematic phase diagram in the $\delta-\epsilon$ plane.

\subsection{Multiscaling}\label{multi}

With the knowledge of the renormalised parameters, and the associated scaling exponents, we now focus on the scaling properties of the structure functions defined above.

\subsubsection{Case I: $g_1^*>0,\,g_2^*=0$}

We consider multiscaling corresponding to the fixed point $g_1^*=\frac{2d}{3}\left(\overline y -d+2\right)>0$ and $g_2^*=0$. At this fixed point, the velocity field decouples from $\phi_\alpha$ in the long wavelength limit. In this case, there is no invariance under a constant shift of all $\phi_\alpha ({\bf x})$; only the rotation in the space of $\phi_\alpha({\bf x})$ keeps the equation of motion invariant. We are interested in the scaling properties of the structure functions $\tilde {\cal S}_{2n}(r)$ defined in (\ref{struct1}) here. %We follow logic outlined above and discuss the details here.

On dimension ground, we can write
\begin{equation}
 \tilde {\cal S}_{2n}(r) \sim r^{2n\chi}\tilde \theta_n(r/\tilde L),
\end{equation}
where, $\tilde \theta_n$ is a dimensionless scaling function of $r/\tilde L$; length scale $\tilde L$ is yet to be specified. Whether $\tilde L$ is a ``large-scale'' (e.g., the integral scale $L$ in a fully developed turbulence), or a ``small scale'' (e.g., the dissipation scale $\eta_d$) is still to be decided. Regardless of the interpretation of $\tilde L$, we assume
\begin{equation}
 \tilde \theta_n\left(\frac{r}{\tilde L}\right) \sim \left(\frac{\tilde L}{r}\right)^{\tilde\Delta_{2n}} \label{gen-exp1}
\end{equation}
in the asymptotic scaling limit~\cite{adjhem,tirtha}. If $\tilde\Delta_{2n}=0$, then scaling function $\theta_n$ approaches a constant in the asymptotic limit, and the scaling of $\tilde {\cal S}_{2n}(r)$ is completely given by $r^{2n\chi}$. On the other hand, a non-zero $\tilde \Delta_n$ implies correction to the scaling given by $r^{2n\chi}$; in fact, a nonlinear dependence of $\Delta_n$ on $n$ should give rise to multiscaling. Notice that if we compare (\ref{gen-exp1}) with the corresponding result (\ref{lin-1}) in the linear theory, we can identify $\tilde L$ with $L$, a large scale. We shall see below that inclusion of nonlinear effects once again identifies $\tilde L$ with $L$; see also Ref.~\cite{tirtha}.

Calculations of $\tilde {\cal S}_{2n}(r)$ involve averaging over spatially non-local quantities with respect to the action functional (\ref{action1}). We use the idea of the operator product expansion (OPE) ~\cite{adjhem,cardy,book} to write
\begin{equation}
 \phi_\alpha^n({\bf x+r})\phi_\alpha^n({\bf x})\sim \sum_m\tilde C_m(r)\tilde {\cal O}_m({\bf x}),\label{ope2-basic}
\end{equation}
in the asymptotic limit of $r/\tilde L\ll 1$. Here, $\tilde C_m(r)$ is a function of the separation $r$ with a power law dependence in the scaling limit, and $\tilde {\cal O}_m({\bf x})$ is an ``operator'' that depends upon $\phi_\alpha({\bf x})$.
Since the rhs of the OPE in (\ref{ope2-basic}) must be invariant under the rotation of $\phi_\alpha({\bf x})$, so should the rhs of (\ref{ope2-basic}) be. The composite operators that can appear on the rhs of (\ref{ope2-basic}) are of the form
\begin{equation}
 \tilde {\cal O}_m({\bf x}) = \left(\phi_\alpha({\bf x})\phi_\alpha({\bf x})\right)^m.
\end{equation}
Evidently, these are invariant under rotation.

\subsubsection{OPE for the linear problem}\label{lin-ope}

The idea of OPE in the problem remains true even in 
the linear limit, for which the fluctuation corrections to all the parameters 
immediately disappear, and exponents $z$ and $\chi$ are given exactly by
(\ref{exact-expo}). We now reconsider the exactly known 
scaling of $\tilde{\cal S}_{2n}(r)$ in the linear case in the context of the OPE 
discussed above and examine the consistency of the latter. Our aim is to obtain (\ref{ans1}) and (\ref{exact12}) by using the prescription of OPE as 
elucidated above: according to that we should have
\begin{equation}
 [\phi_\alpha({\bf x}+{\bf r})\phi_\alpha ({\bf x})] \sim \tilde {\cal C}_0(r){\bf I} + 
\tilde{\cal C}_2(r) [\phi_\alpha ({\bf x})]^2, \label{ope2}
\end{equation}
such that
\begin{equation}
 {\cal \tilde S}_2(r)\sim {\cal \tilde C}_0(r){\bf I} + 
{\cal \tilde C}_2(r) \langle[\phi_\alpha ({\bf x})]^2\rangle.
\end{equation}
Here, $\bf I$ is the identity operator.
{ Now in the linear theory for $\overline y + 2<d$, or $\chi <0$, the left had side of (\ref{ope2}) scales as $b^{2\chi}$, when all spatial coordinates are scaled by $b$.  Each term on the rhs of (\ref{ope2}) on dimensional ground must scale the same way under spatial rescaling as the lhs.  This consideration gives   
\begin{equation}
 \tilde {\cal C}_0(r)\sim r^{2\chi}.
\end{equation}
On the other hand, 
\begin{equation}
 \langle [\phi_\alpha ({\bf x})]^2\rangle \sim a_0^{2\chi}
\end{equation}
where $a_0$ is the small scale cut-off. This gives on the dimensional ground
\begin{equation}
 \tilde {\cal C}_2(r)\sim r^0.
\end{equation}
Therefore, in the scaling limit of $r\gg a_0$, the first term on the rhs of (\ref{ope2}) dominates. This gives
\begin{equation}
 \tilde{\cal S}_{2}(r)\sim r^{2\chi}\sim r^{-(2-d+\overline y)}.
\end{equation}
Using the linearity of the $\phi_\alpha$-dynamics then,
\begin{equation}
 \tilde{\cal S}_{2n}(r)\sim r^{2\chi}\sim r^{-n(2-d+\overline y)},
\end{equation}
in agreement with (\ref{exact11}).}

{ In contrast, for $\overline y+2>d$ or $\chi >0$  the 
second term in the right hand side
of (\ref{ope2}) dominates. This may be argued as follows. On dimensional ground, we still have $\tilde{\cal C}_2(r)\sim r^{2\chi}$. On the other hand, with $\chi>0$
\begin{equation}\label{Ly}
\langle [\phi_\alpha ({\bf x})]^2\rangle \sim
L^{\overline y+2-d},
\end{equation}
{ an $L$-dependence identical to that in (\ref{ans1})}, giving 
$\tilde {\cal C}_2(r)\sim r^0$, again using dimensional analysis. } Therefore, in the scaling limit $r\ll L$, the second term on the rhs of (\ref{ope2}) dominates over the first term. We thus conclude from
{ Eq.~(\ref{Ly}) that }
\begin{equation}
 \tilde {\cal S}_2(r) \sim 
L^{\overline y+2-d},
\end{equation}
{unsurprisingly same as (\ref{ans1}).} This may be extended to higher 
order structure ($n>1$) 
easily. For instance for $n=2$, the most dominant operator { that contributes 
to}
$\tilde{\cal S}_4(r)$ { in the scaling limit} is 
$( \phi_\alpha({\bf x}))^4$. It is easy to see $\langle (
\phi_\alpha({\bf x}))^4\rangle \sim L^{2(\overline y-d)}$, giving $\tilde{\cal C}_4 
(r)\sim r^4$. Putting together everything then, 
\begin{equation}
\tilde {\cal S}_4(r)\sim 
L^{2\overline y +4-2d}, 
\end{equation}
which is same as that obtained by direct calculations above; see (\ref{exact12}) above. 
This evidently lends credence to our analysis even when 
$\lambda\neq 0$. In general, we conclude
\begin{equation}
 \tilde {\cal S}_n(r)\sim 
L^{n(\overline y +2-d)}.
\end{equation}
We note that only when $\chi>0$, the structure functions can depend on $L$. 
Having re-established the exactly known scaling exponents of $\tilde {\cal 
S}_{2n}(r)$ in the linear limit by the arguments of OPE, we now analyse the 
nonlinear cases below. 

\subsubsection{OPE for Model I}\label{OPE-I}

We now use the idea of OPE to calculate the scaling of ${\cal \tilde S}_{2n}(r)$ in the nonlinear case: We start from the general expansion of OPE
\begin{equation}
 {\cal \tilde S}_{2n}(r)\sim \sum_{m}\tilde C_m(r)\langle {\cal \tilde O}({\bf x})\rangle. \label{strut-ope}
\end{equation}
Thus, in order to have  a one-loop 
renormalised theory for multiscaling we must now find out how the na\"ive 
scaling of ${\cal \tilde O}_m({\bf x})$ changes due to fluctuations. We continue to consider the case with $\chi>0$ for which composite operators are expected to contribute to the scaling properties of the structure functions. We are required to find out which composite operator on the rhs of (\ref{strut-ope}) makes the most dominant contribution in the asymptotic limit $L\gg r$. Na\"ive expansion of the lhs of (\ref{strut-ope}) suggests that $\tilde {\cal O}_m({\bf x})$ with $m=n$ should make the desired most dominant contribution. 
%This will allow 
%us to determine the most dominant term in (\ref{compo1}) within a one-loop 
%renormalised theory.
Assuming that $\tilde{\cal O}_n({\bf x})$ do not generate any higher order relevant operators,  we must
now  
calculate the one loop renormalisation of ${\cal\tilde O}_n({\bf x})$ for 
arbitrary $n$ and then 
find the scaling forms
for the renormalised composite operators; see Appendix~\ref{Omx} for the relevant 
one-loop 
Feynman 
diagrams { contributing to the} renormalisation of ${\cal \tilde O}_m({\bf 
x})$;  see also Appendix~\ref{higher-op} for some related technical discussions (including higher order operators). 
We find {
\begin{equation}
 \langle\tilde{\cal  O}^<_m({\bf x})\rangle=\langle\tilde {\cal O}_m({\bf 
x})\rangle\left[1+\delta m'\right],
\end{equation}
where, $\delta m'$ is the fluctuation correction.

We now consider the the one-loop correction for $\tilde {\cal O}_n({\bf x})$. The one-loop diagram is shown in Fig.~\ref{one-comp} in Appendix~\ref{Omx}.

Evaluation of this diagram is shown in Appendix. We find
\begin{equation}
 \tilde {\cal O}^<_n({\bf x})=\tilde {\cal O}_n({\bf x})\left[1+2\frac{D_1g^2}{2\nu^3}\int_{\Lambda/b}^\Lambda\frac{d^dq}{(2\pi)^d}\frac{|q|^{-y}}{q^2}\right].
\end{equation}
This results into the RG flow equation for $\langle \tilde {\cal O}^<_n({\bf x})\rangle$
\begin{equation}
 \frac{d\langle\tilde {\cal O}_n\rangle}{dl}=\langle\tilde {\cal O}_n\rangle \left[2n\chi +\tilde \delta_n\right],
\end{equation}
where $\tilde\delta_n = \frac{nD_1g^2}{\nu^3}=ng_1$ is {\em linear} in $n$.
This implies
\begin{equation}
 \langle\tilde {\cal O}_n({\bf x})\rangle\sim L^{2n\chi +\tilde \delta_n}\sim L^{\tilde\Delta n}.
\end{equation}
This gives for the scaling behaviour of the structure functions $\tilde {\cal S}_{2n}(r)$ as
\begin{equation}
 \tilde {\cal S}_{2n}(r)\sim r^{2n\chi} f_{2n}\left(\frac{L}{r}\right)\sim r^{2n\chi}\left( \frac{L}{r}\right)^{\tilde\Delta_n}\sim r^{-\tilde\delta_n}L^{\tilde\Delta_n},
\end{equation}
where $\tilde\Delta_n$ or $\tilde \delta_n$ should be evaluated at the RG fixed point with $g_1=g_1^*$.
In this linear theory, $\tilde {\cal S}_{2n}(r)$ is independent of $r$ in the asymptotic limit $r\ll L$, but depend on $L$ with exponents that vary linearly with $n$. Nonlinear effects change this picture. Structure functions $\tilde {\cal S}_{2n}(r)$ now do depend on $r$, but with {\em negative} exponents, i.e., they decrease with $r$:
\begin{equation}
 \tilde \zeta_n=-\tilde \delta_n = -ng_1^*=n\frac{2d}{3}\delta.
\end{equation}
On the other hand, their dependence on $L$ are now more sensitive; the exponents that characterise their dependence on $L$ are {\em positive} and {\em bigger} than their values in the linear theory. We notice that the scaling exponents $\tilde\zeta_n$ are ${\cal O}(\delta)$ at the lowest order in the perturbation theory, consistent with the structure of the perturbation theory with $\delta$ and $\epsilon$ as ``small expansion parameters''.
%Contrast this with the results for Case I (with $g_2^*=0$ and $g_1^*=..)$, where the relevant structure functions ${\cal S}_{2n}(r)$ {\em grow} with $r$, albeit slower than the prediction from the linear theory.

\subsubsection{Case II: $g_1^*=0,\,g_2^*>0$}

We now consider the part of the phase space in the $y-\overline y$ plane where $g^*_2=d(y-d)/(d-1)>0$ and $g^*_1=0$. In this part of the phase space, the mode coupling term is {\em irrelevant} (in a RG sense), and thus there is an emergent symmetry under $\phi_\alpha({\bf x},t)\rightarrow \phi_\alpha({\bf x},t) + {\rm const.}$ in the long wavelength limit. The effective model for each of $\phi_\alpha$ is in fact identical to that used for the passive scalar turbulence problem~\cite{tirtha}. In this case, it is meaningful to consider the scaling properties of the structure functions ${\cal S}_{2n}(r)$ defined above. We revisit the calculation for the scaling properties of ${\cal S}_{2n}(r)$ for the sake of completeness below. On purely dimensional ground, we expect
\begin{equation}
 {\cal S}_{2n}(r)=r^{2n\chi} \theta_n(r/\tilde L),
\end{equation}
where $\theta_n$ is a dimensionless scaling function of $r/\tilde L$, $\tilde L$ is a length scale that is yet to be specified. %Where $\tilde L$ is a ``large-scale'' (e.g., the integral scale $L$ in a fully developed turbulence), or a ``small scale'' (e.g., the dissipation scale $\eta_d$) remains to be seen. 
As before, independent of the interpretation of $\tilde L$, we write
\begin{equation}
 \theta_n\left(\frac{r}{\tilde L}\right) \sim \left(\frac{\tilde L}{r}\right)^{\Delta_{2n}} \label{gen-exp}
\end{equation}
in the asymptotic scaling regime~\cite{tirtha}. For $\Delta_{2n}=0$, the scaling function $\theta_n$ approaches a constant in the asymptotic limit, and  ${\cal S}_{2n}(r)\sim r^{2n\chi}$. In contrast, a non-zero $\Delta_n$ implies correction to the scaling given by $r^{2n\chi}$.%; in fact, a nonlinear dependence of $\Delta_n$ on $n$ may give rise to multiscaling. Comparing (\ref{gen-exp}) with the corresponding result (\ref{lin-1}) in the linear theory, we  identify $\tilde L$ with $L$, a large scale~\cite{tirtha}.%. We shall see below that inclusion of nonlinear effects once again identifies $\tilde L$ with $L$; see also Ref.~\cite{tirtha}.

We briefly revisit the results and the discussions in Ref.~\cite{tirtha}. We express the structure functions in terms of {\em local composite operators}
\begin{equation}
 [\phi_\alpha({\bf x+r}-\phi_\alpha({\bf x})]^{2n}\sim \sum_m C_m(r){\cal O}_m({\bf x}).\label{ope1}
\end{equation}
Using (\ref{ope1}),
\begin{equation}
 {\cal S}_{2n}(r)\sim C_n(r)\langle {\cal O}_n({\bf x})\rangle.
\end{equation}
In the asymptotic limit of $L/r\rightarrow \infty$, we have
\begin{equation}
 {\cal S}_{2n}(r)\sim r^{2n\chi}\left(\frac{L}{r}\right)^{\Delta_n}.
\end{equation}

As explained in Ref.~\cite{tirtha}, the most dominant operator is
\begin{equation}
 {\cal O}_n({\bf x})=(\partial_i \phi_\alpha({\bf x})\partial_i\phi_\alpha({\bf x})^n.
\end{equation}
As discussed in Ref.~\cite{tirtha} in details, there are one-loop corrections to ${\cal O}_n({\bf x})$ with the fluctuation-corrected composite operator
\begin{equation}
 {\cal O}_n^<({\bf x})={\cal O}_n({\bf x})[1+\delta n],
\end{equation}
where 
\begin{equation}
 \delta n = \frac{\lambda_1^2 Dn(d-1)(d+2n)}{\nu d(d+2)},
\end{equation}
giving for the renormalised composite operator
\begin{equation}
\langle {\cal O}_n({\bf x})\rangle \sim L^{2n(\chi-1)+\delta n}. 
\end{equation}
This implies
\begin{equation}
 {\cal S}_{2n}(r)\sim r^{2n-\delta n}\sim r^{\zeta_n},
\end{equation}
where
\begin{equation}
 \zeta_n=n(\overline y -y)+\frac{(y-d)n(d+2n)}{d+2}.
\end{equation}
Thus, $\zeta_n$ depends nonlinearly on $n$, a hallmark of multiscaling. Structure functions ${\cal S}_{2n}(r)$ grows with $r$ in the renormalised theory, albeit slower than in the linear theory. Contrast this with the results for Case I (i.e., with $g_1^*>0,\,g_2^*=0$). In that case, the relevant structure functions ${\cal \tilde S}_{2n}(r)$ {\em decay} with $r$ in the renormalised theory. Here too, $\delta_n\sim {\cal O}(\epsilon)$.

\section{Summary and outlook}\label{summ}

In this work, we have studied the issue of multiscaling in stochastically driven nonlinear dynamical models. To that end, we have considered a simple, conceptual model in the form of the conserved dynamics of classical Heisenberg spins above $T_c$ that included a mode-coupling contribution, advected by a stochastic velocity field  that is assumed to be Gaussian-distributed and independent of the spin dynamics, and an additive noise. We define appropriate structure functions ${\cal \tilde S}_{2n}(r)$ that should capture the universal scaling when the mode coupling term is relevant in a RG sense. When it is irrelevant, a different set of structure functions ${\cal S}_{2n}(r)$ that are essentially same as those used for the passive scalar turbulence problem is useful for this purpose. We show that when the velocity and additive noise correlations are sufficiently long-ranged, as measured by the spatial scaling of the their variances, the structure functions start to depend on the system size $L$. The precise $L$-dependences of these two sets of the structure functions are, however, quite different. More interestingly, ${\cal \tilde S}_{2n}(r)$ {\em do not} show any multiscaling, i.e., the associated scaling exponents ${\tilde \zeta}_{2n}$ are {\em linear} functions of $n$. In contrast, ${\cal S}_{2n}$ show genuine multiscaling identical to those found in Refs.~\cite{adjhem,kupi,tirtha}. We also show in Appendix~\ref{model-II} that these results are rather insensitive to the specific choice for the large damping limit of the the flow field equation, so long as the flow field remains autonomous. On the other hand, nonlinear effects can generally be very crucial in controlling the nature of multiscaling. In fact, in this particular case, the nonlinear mode coupling term, when relevant, can lead to simple scaling instead of multiscaling.

Our calculational scheme is technically challenging. Due to the mode coupling term, composite operators of the spin variables of a particular order are now connected to all higher order composite operators. We have argued in Appendix~\ref{higher-op} that due to the conservation law nature of the spin dynamics, the perturbatively generated higher order composite operators from a composite operator of a given order should be subleading due to the appearance of additional gradient operators. 

The quantitative accuracy of our results are limited by the low order of the perturbation theory. Nonetheless, we still expect that the qualitative features of our results will hold even in a more sophisticated perturbation theory, or in numerical solutions of the equations of motion. It will indeed be interesting to test our results by numerically solving the governing equations of motion. In the more realistic problem multiscaling in a system with coupled variables, e.g., $3d$ magnetohydrodynamic turbulence or binary fluid turbulence, both the velocity field and the second dynamical field, e.g., the magnetic field or the concentration field display non-trivial multiscaling. While our results cannot be obviously carried over to these problems due to the simplified and admittedly artificial nature of our model equation and the significant technical challenges involved in $3d$ magnetohydrodynamic or binary fluid turbulence, we can certainly conclude with a good degree of confidence that the presence of the different nonlinear terms can affect the detailed form of the multiscaling exponents in a nontrivial manner~\cite{abmhd,binfluid}. In this work, we have been concerned with the multiscaling (or scaling) of the (appropriately defined) equal-time structure functions. More recently, the concepts of scaling and multiscaling have been extended to their dynamic analogues as well~\cite{dyn-mult}. It will be interesting to study the effects of the nonlinear term in the model equation studied here on possible dynamic multiscaling. Recent studies have indicated that the anomalous scaling in the Kraichnan model actually resembles more like that for the Burgers equation for pressureless turbulence~\cite{sreeni2}, instead of the form predicted by perturbation theories. Similar studies could be undertaken for the present nonlinear model to find out the validity of the perturbation theory results discussed in the work. We hope in future our studies here will provide new impetus to future analytical work to unearth the physics of multiscaling in forced hydrodynamic turbulence and its analogous systems.

\appendix

%\section{Bare propagators and correlators}

\section{One-loop corrections to $\nu$ for $\overline y>-2$}\label{feyn-nu}

We start with the 
 bare propagators and correlators in the action (\ref{action1}), which are given by

\begin{eqnarray}
 \langle \hat\phi_\alpha ({\bf q},\omega) \hat\phi_\beta (-{\bf q},-\omega) \rangle &=& 
0\\ \nonumber
 \langle \hat\phi_\alpha ({\bf q},\omega) \phi_\beta (-{\bf q},-\omega) \rangle &=& 
\frac{\delta_{\alpha\beta}}{i\omega+\nu q^2}\\ \nonumber
 \langle \hat\phi_\alpha (-{\bf q},-\omega) \phi_\beta ({\bf q},\omega) \rangle &=& 
\frac{\delta_{\alpha\beta}}{-i\omega+\nu q^2}\\ \nonumber
 \langle \phi_\alpha ({\bf q},\omega) \phi_\beta (-{\bf q},-\omega) \rangle &=& \frac{2 \delta_{\alpha\beta}
\overline D 
q^{-\overline y}}{\omega^2+\nu^2 q^4}.
\end{eqnarray}

\begin{widetext}

\begin{figure}[htb]
 \includegraphics[width=7cm]{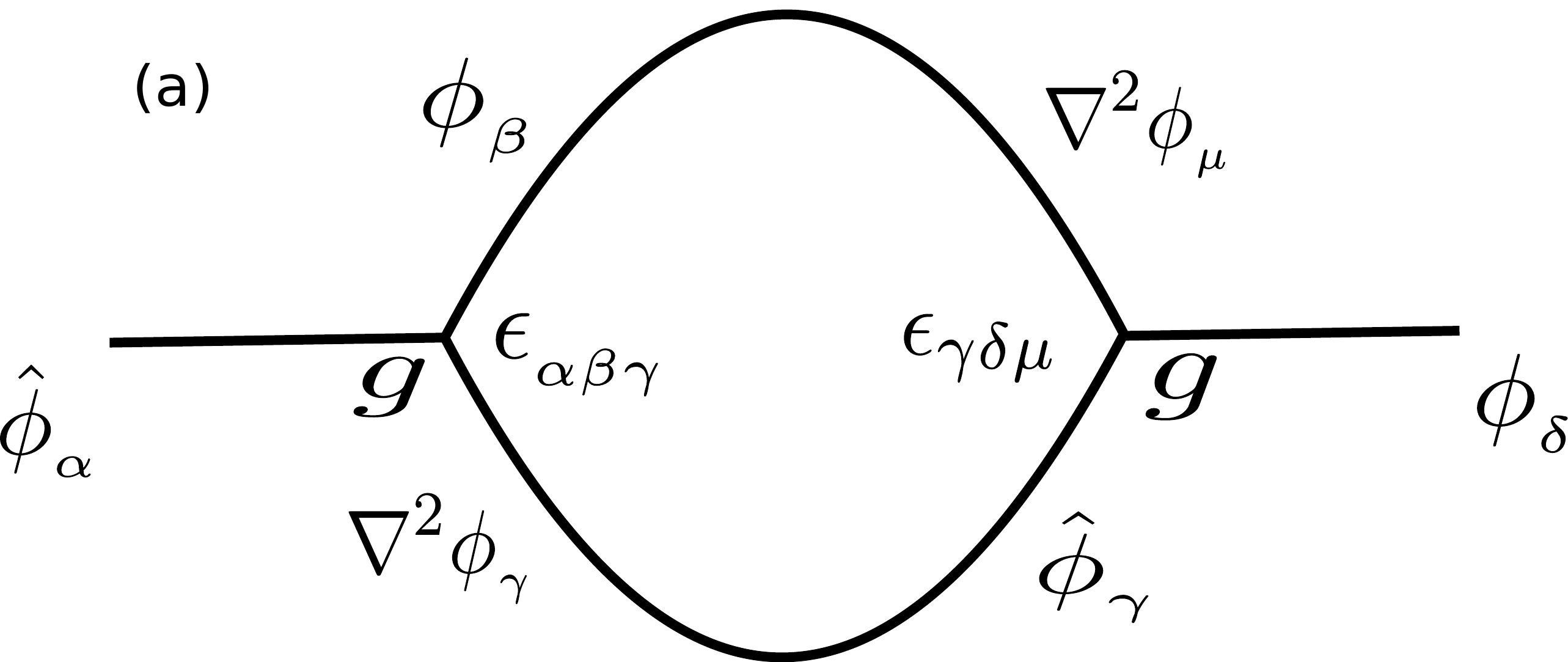}\hfill
 \includegraphics[width=7cm]{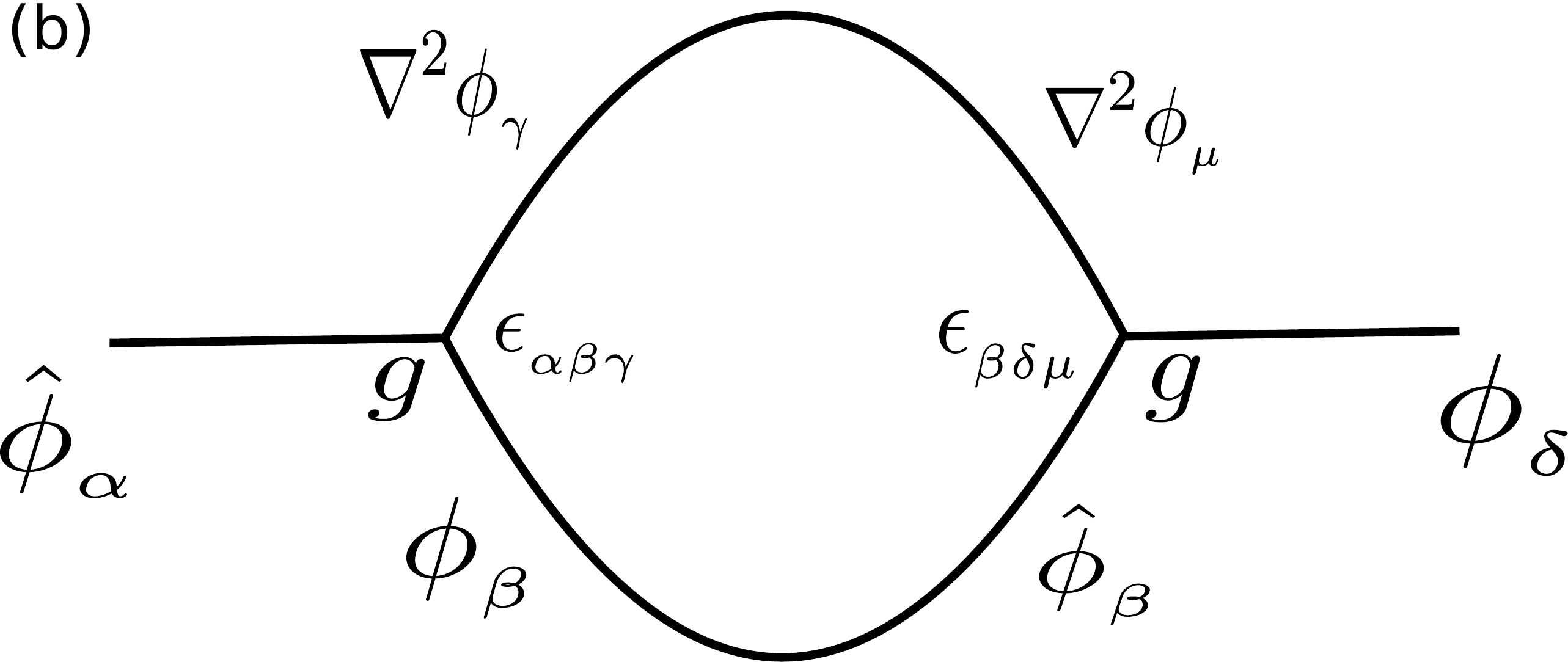}\\
 \includegraphics[width=7cm]{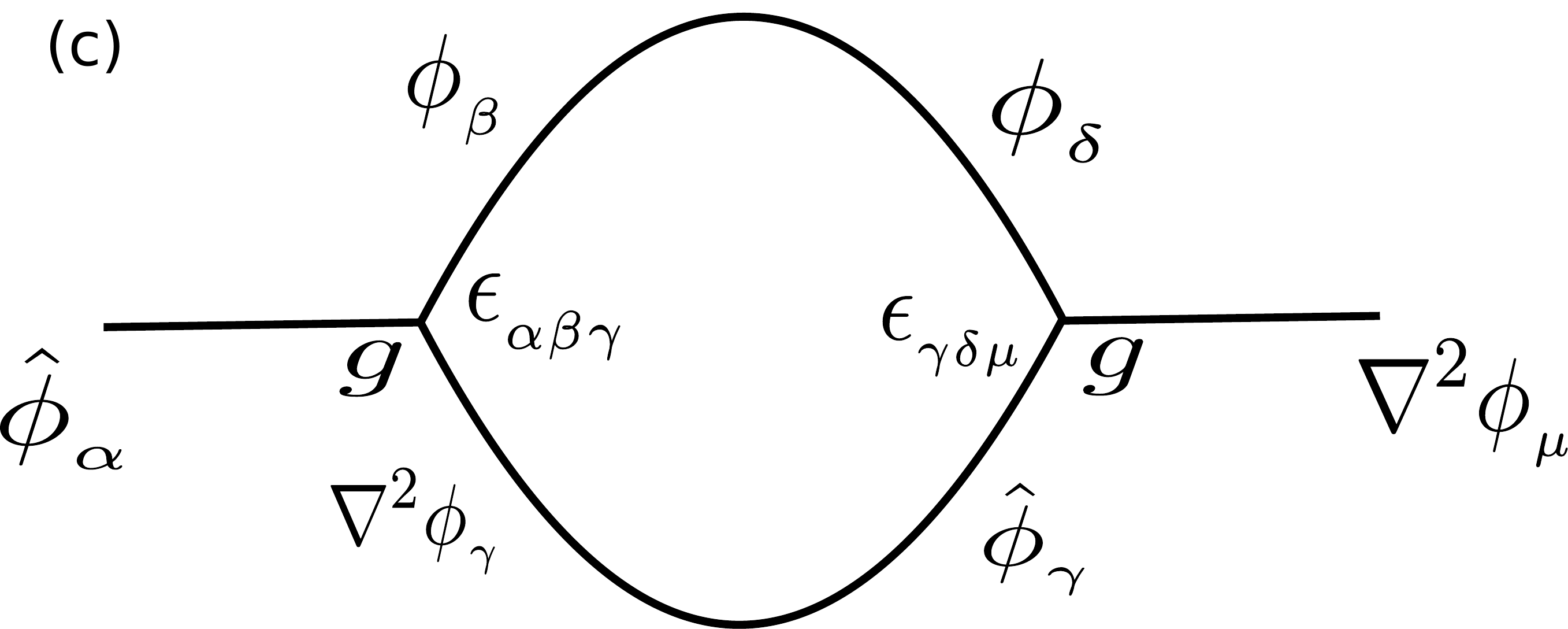}\hfill
 \includegraphics[width=7cm]{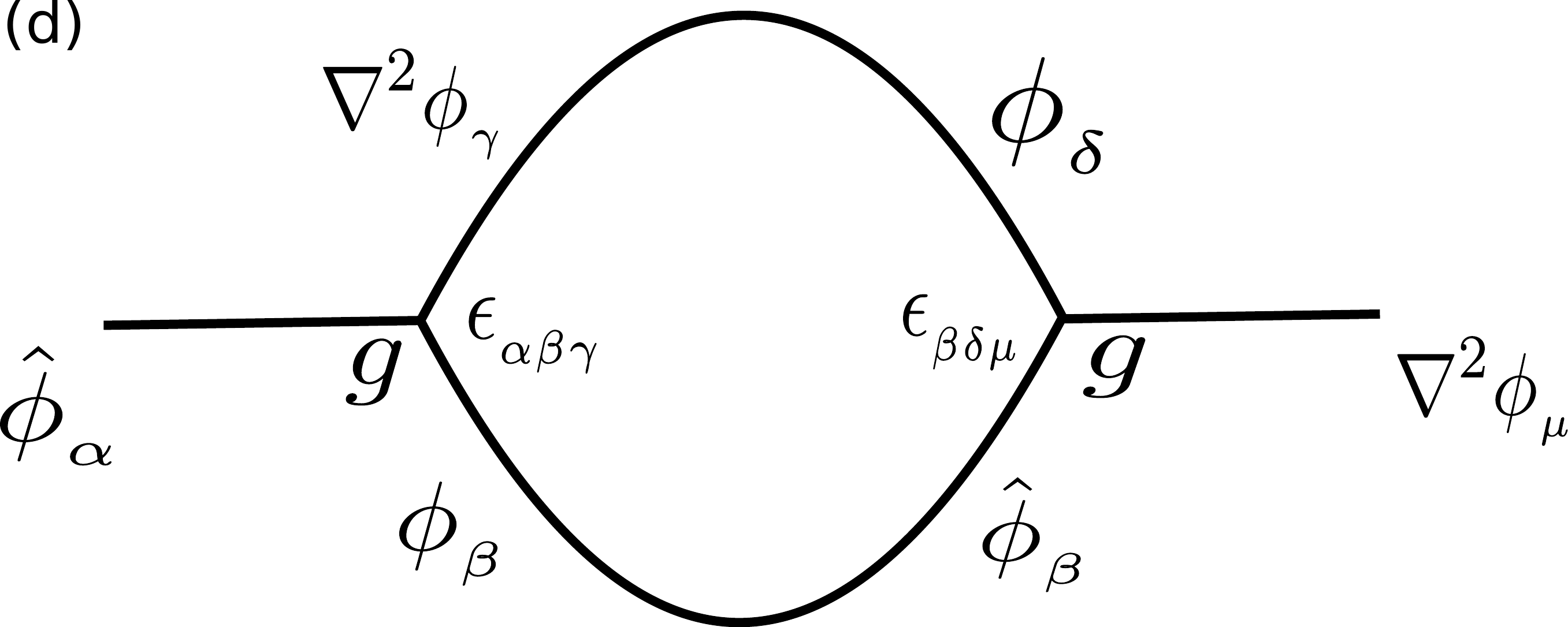}\\
 \includegraphics[width=7cm]{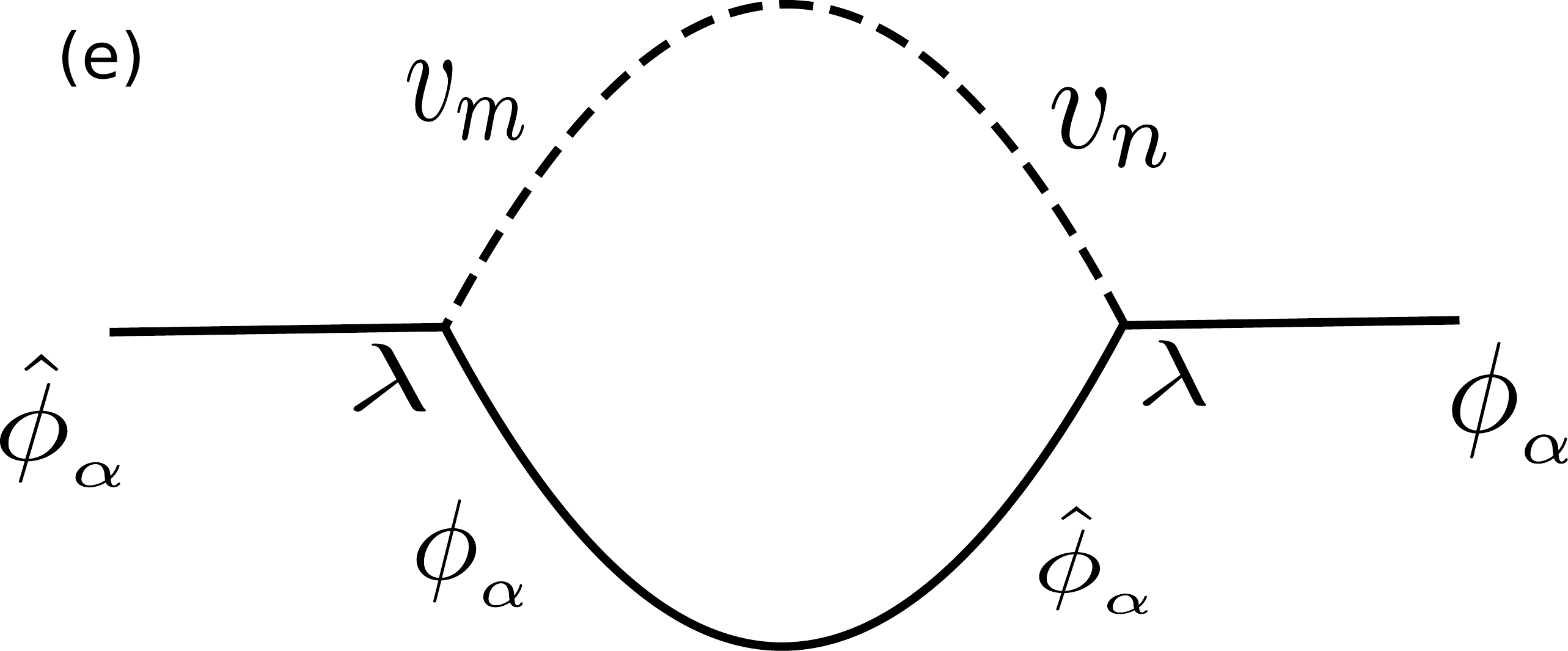}
 \caption{One-loop Feynman graphs that correct $\nu$. Diagrams (a-d) have their origin in the mode-coupling term, where as diagram (e) comes from the advective nonlinear term.} 
\label{nu-diag}
\end{figure}

\end{widetext}

The relevant Feynman diagrams are shown in Fig.~\ref{nu-diag}. We discuss the diagrams which originate from the mode-coupling vertex in detail below.

(i) Diagrams (\ref{nu-diag})(a) and (b) combine to yield (each has symmetry factor 2)

\begin{widetext}
 
\begin{eqnarray}
 &&g^2\hat\phi_\alpha\phi_\delta\epsilon_{\alpha\beta\gamma}\int\frac{d^dq}{(2\pi)^d}\frac{d\Omega}{2\pi}\frac{2\overline Dq^{-\overline y}}{\Omega^2 +\nu^2 q^4}\frac{1}{i\Omega + \nu({\bf k-q})^2}\left(\epsilon_{\gamma\delta\mu}\delta_{\beta\mu}q^2({\bf k-q})^2 + \epsilon_{\beta\delta\mu}\delta_{\gamma\mu}q^4\right)\nonumber \\
 &&=g^2\hat\phi_\alpha\phi_\delta\epsilon_{\alpha\beta\gamma}\frac{\overline D}{2\nu^2}\int\frac{d^dq}{(2\pi)^d}\frac{q^{-\overline y}}{q^2 + ({\bf k-q})^2}\left(\epsilon_{\gamma\delta\mu}\delta_{\beta\mu}q^2({\bf k-q})^2 + \epsilon_{\beta\delta\mu}\delta_{\gamma\mu}q^4\right).
\end{eqnarray}

%\end{widetext}

 We symmetrise this to obtain
 
%\begin{widetext}
 
 \begin{eqnarray}
  &&g^2\hat\phi_\alpha\phi_\delta\epsilon_{\alpha\beta\gamma}\epsilon_{\gamma\delta\beta}\frac{\overline D}{4\nu^2}\int\frac{d^dq}{(2\pi)^d}\frac{({\bf k-q})^2 - q^2}{q^2 + ({\bf k-q})^2}\left[q^{-\overline y} - |{\bf k-q}|^{-\overline y}\right]\nonumber \\&&= g^2\hat\phi_\alpha\phi_\delta\epsilon_{\alpha\beta\gamma}\epsilon_{\gamma\delta\beta}\frac{\overline D}{4\nu^2}\int\frac{d^dq}{(2\pi)^d}\frac{\bf k\cdot q}{q^2}\left[q^{-\overline y}-q^{-\overline y}\left(1+\overline y \frac{\bf k\cdot q}{q^2}\right)\right]\nonumber \\&& = -g^2\hat\phi_\alpha\phi_\delta\left(\delta_{\alpha\delta}\delta_{\beta\beta}-\delta_{\alpha\beta}\delta_{\beta\delta}\right) \frac{\overline D}{4\nu^2}k_dk^2\int dq q^{d-\overline y -3}=-\delta_{\alpha\delta}\frac{\overline y\overline D}{2d\nu^2}k_dk^2\int dq q^{d-\overline y -3}\hat\phi_\alpha\phi_\delta.
 \end{eqnarray}

\end{widetext}

(ii) Diagrams (\ref{nu-diag})(c) and (d) combine to yield (each has a symmetry factor 2)

\begin{widetext}
 
\begin{eqnarray}
 &&\hat\phi_\alpha\phi_\mu g^2\epsilon_{\alpha\beta\gamma}k^2\int\frac{d^dq}{(2\pi)^d}\frac{d\Omega}{2\pi} \frac{2\overline D |q|^{-\overline y}}{\Omega^2 +\nu^2 q^4}\frac{1}{i\Omega +\nu({\bf k-q})^2}\left(\epsilon_{\gamma\delta\mu}\delta_{\beta\mu}({\bf k-q})^2 + \epsilon_{\beta\delta\mu}\delta_{\delta\mu}q^2\right)\nonumber \\&&= \hat\phi_\alpha\phi_\mu g^2\left(\epsilon_{\alpha\beta\gamma} \epsilon_{\gamma\delta\mu}\delta_{\beta\delta}+\epsilon_{\alpha\gamma\beta}\epsilon_{\gamma\delta\mu}\delta_{\beta\delta}\right)\frac{\overline Dk^2}{2\nu^2}\int\frac{d^d}{(2\pi)^d}\frac{1}{q^{\overline y+2}}=0.
\end{eqnarray}

\end{widetext}

The contribution from diagram~\ref{nu-diag}(e) reads

\begin{equation}
 -\hat\phi_\alpha \phi_\alpha\lambda^2 D_1\frac{d-1}{d}k_d \int dq q^{d-1-y}.
\end{equation}

\section{Composite operators}\label{Omx}

\begin{figure}[htb]
\includegraphics[width=7cm]{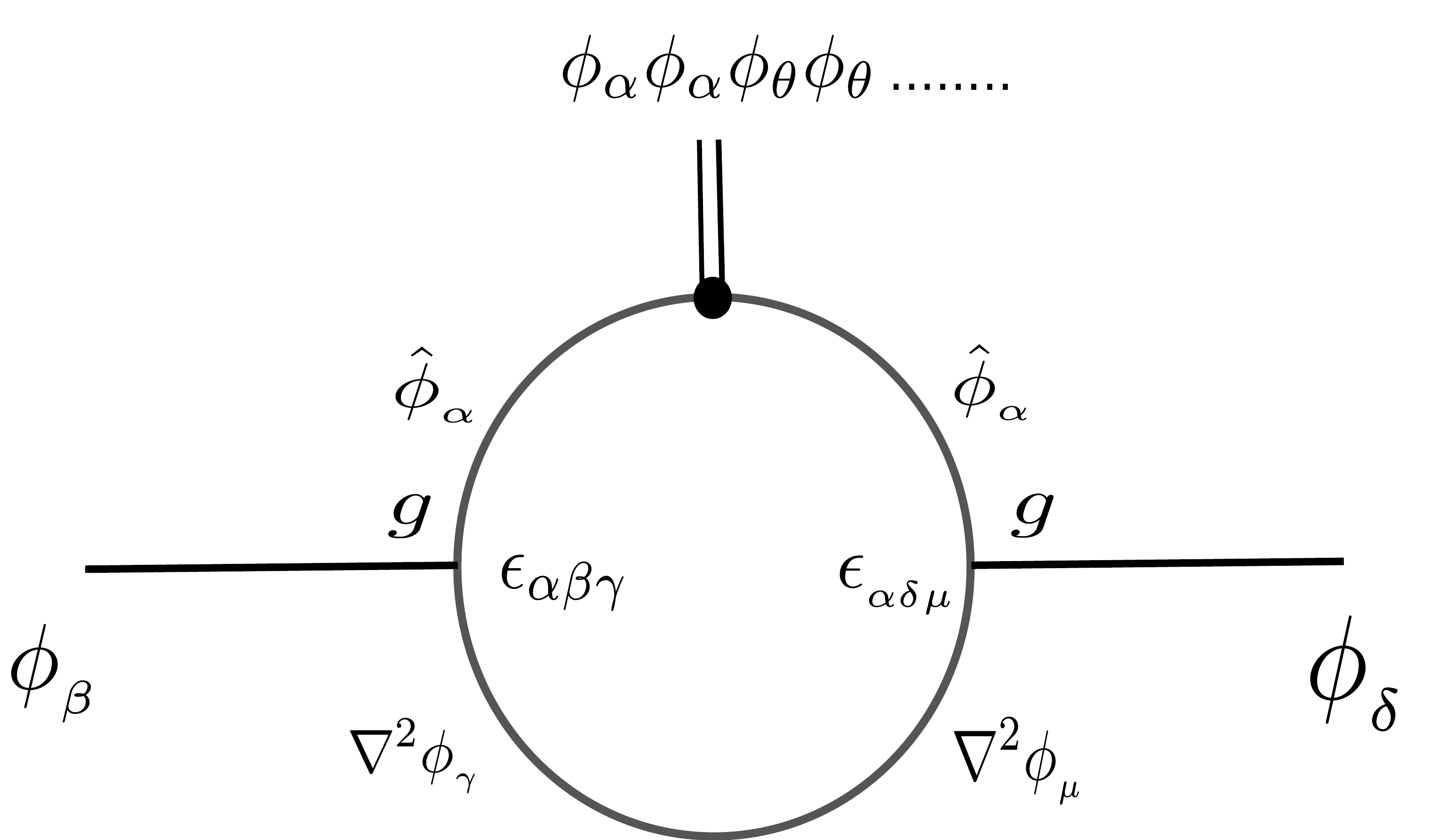}\\
\includegraphics[width=7cm]{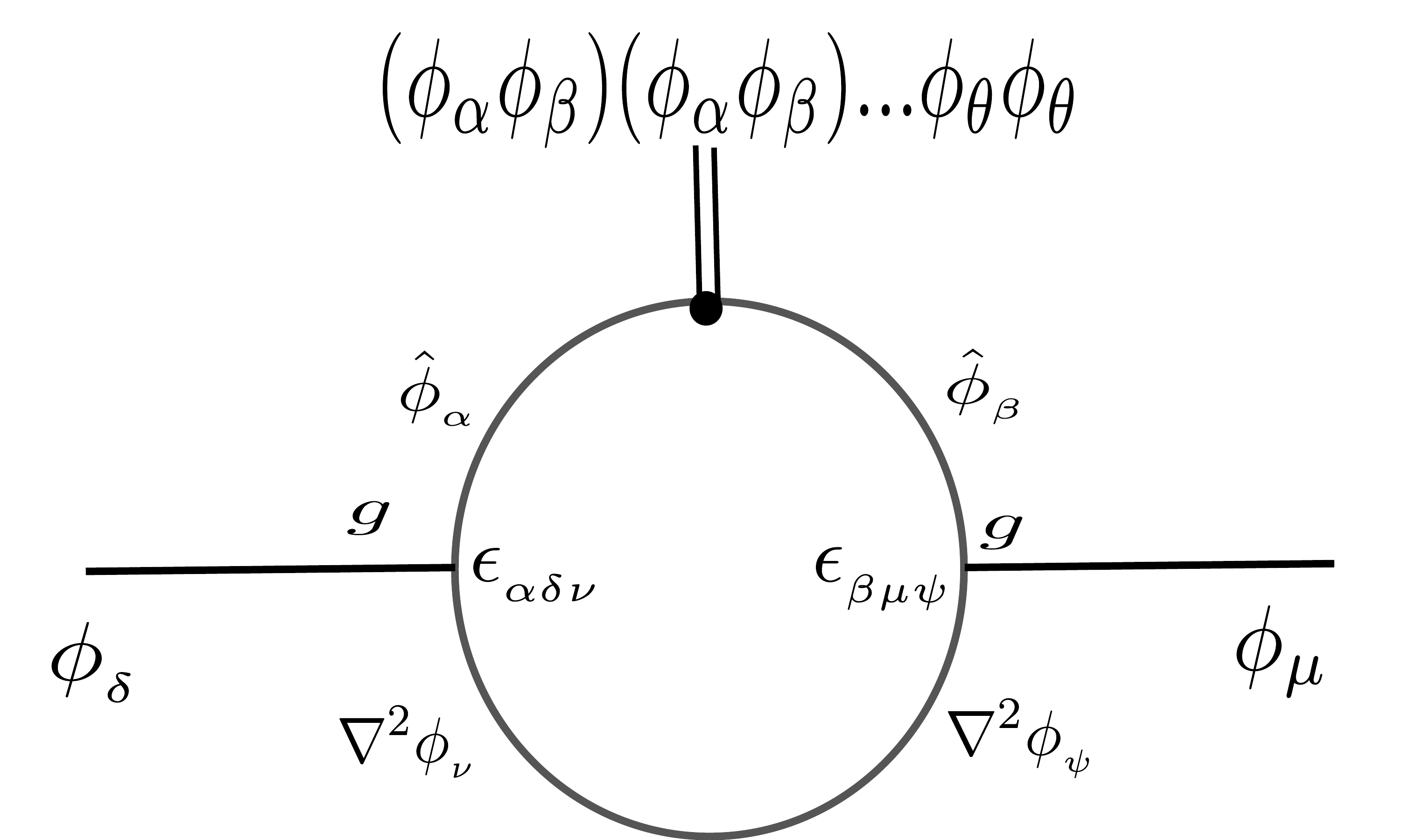}
 \caption{One-loop Feynman graphs for the composite operator $\tilde {\cal O}_n({\bf x})$. In both these diagrams, a small thick circle together with the pair of small vertical parallel lines represent a composite operator, that is expanded to form the one-loop diagrams in the lowest order perturbation theory (see text).} 
\label{one-comp}
\end{figure}

We calculate the one-loop corrections to the composite operators $\tilde {\cal O}_n({\bf x})$. The only relevant contribution comes from the model coupling vertex in (\ref{action1}), since the advective vertex is ${\cal O}(q)$. The one-loop diagram in Fig.~(\ref{one-comp}) may be constructed by contracting any two of the $\phi_\alpha$-fields in ${\cal \tilde O}_n$ with the two trilinear anhamornic terms, each of the form $(\nabla_m\hat\phi_\alpha)\epsilon_{\alpha\beta\gamma}\phi_\beta\nabla_m\phi_\gamma$. There are two distinct ways these contractions can be done:

(i) The two $\phi$-fields in ${\cal \tilde O}_n$ which are to be contracted with the anhamornic terms have the same indices. The contribution reads
\begin{equation}
 n\frac{\overline Dg^2}{\nu^3}\phi_\alpha\phi_\alpha\int\frac{d^dq}{(2\pi)^d}\frac{1}{q^{\overline y+2}}.
\end{equation}

(ii) The two $\phi$-fields in ${\cal \tilde O}_n$ which are to be contracted with the anhamornic terms have different indices. The corresponding contribution reads

\begin{widetext}
 
\begin{eqnarray}
 &&2n(n-1)\phi_\beta\phi_\lambda g^2\epsilon_{\alpha\beta\gamma}\epsilon{\mu\lambda\nu}\int\frac{d^dq}{(2\pi)^d}\frac{d\Omega}{2\pi} \frac{2\overline Dq^{4-\overline y}\delta_{\gamma\nu}}{(\Omega^2 +\nu^2 q^4)^2}=n(n-1)\overline D \phi_\beta\phi_\lambda g^2\epsilon_{\alpha\beta\gamma}\epsilon{\mu\lambda\gamma}\int\frac{d^dq}{(2\pi)^d}\frac{1}{q^{\overline y+2}}\nonumber \\
 &=&n(n-1)\overline D\phi_\beta\phi_\lambda g^2[\delta_{\alpha\mu}\delta_{\beta\lambda}-\delta_{\alpha\lambda}\delta_{\beta\mu}]\phi_{\beta\lambda}=0.
\end{eqnarray}

\end{widetext}
Thus at the one-loop order, the fluctuation corrections to ${\cal \tilde O}_n$ are linear in $n$.

\section{Higher order composite operators}\label{higher-op}

Due to the mode-coupling anharmonic term in (\ref{action1}), a composite operator of the form ${\tilde O}_m({\bf x})= (\phi_\alpha({\bf x})\phi_\alpha({\bf x}))^m$ for some positive  integer $m$ can perturbatively generate a composite operator ${\tilde O}_p({\bf x})$, where $p>m$. To the lowest order perturbation expansion, $p=m+1$; see Fig.~\ref{higher}. 
\begin{figure}[htb]
\includegraphics[width=6cm]{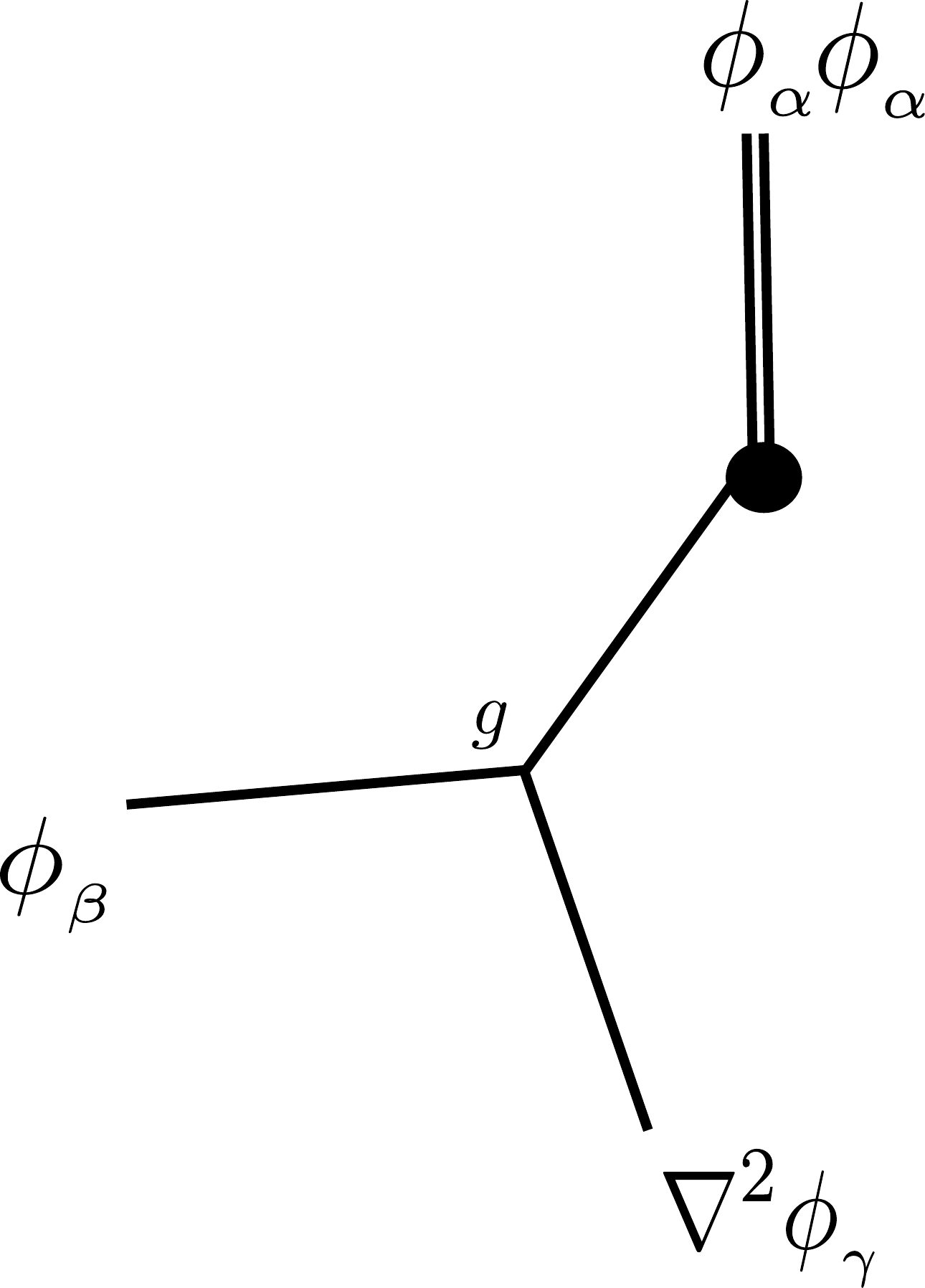}
 \caption{Generation of ${\cal \tilde O}_3({\bf x})$ from ${\cal \tilde O}_2({\bf x})$ in the lowest order perturbation theory.}\label{higher}
\end{figure}

This is distinct from the traditional passive scalar problem, where due to the linearity of the equation of motion, there is no possibility of generating higher order composite operators. This unambiguously allowed one to determine the operator that made the largest contribution to the multiscaling of the structure functions in the passive scalar problem. This holds true in the present problem in the subspace where the renormalised coupling of the nonlinear mode coupling term vanishes. In the other region of the phase space, where this coupling is relevant, this is generally not true. We however note that \\\\
(i) $\langle {\cal \tilde O}_3({\bf x})\rangle$ vanishes due to the rotational invariance in the order parameter space. Thus, in the lowest order perturbative expansion $\langle{\cal \tilde O}_2({\bf x})\rangle$ {\em does not} connect to any higher order composite operators.\\\\
(ii) At higher order perturbation expansions, ${\cal \tilde O}_2({\bf x})$ can however generate higher order composite operators. For instance, in the second order in $g$ (the coupling constant for the mode coupling nonlinearity), ${\cal \tilde O}_2({\bf x})$ can generate ${\cal \tilde O}_4({\bf x})$; see Fig.~\ref{higher1}.
\begin{figure}[htb]
\includegraphics[width=6cm]{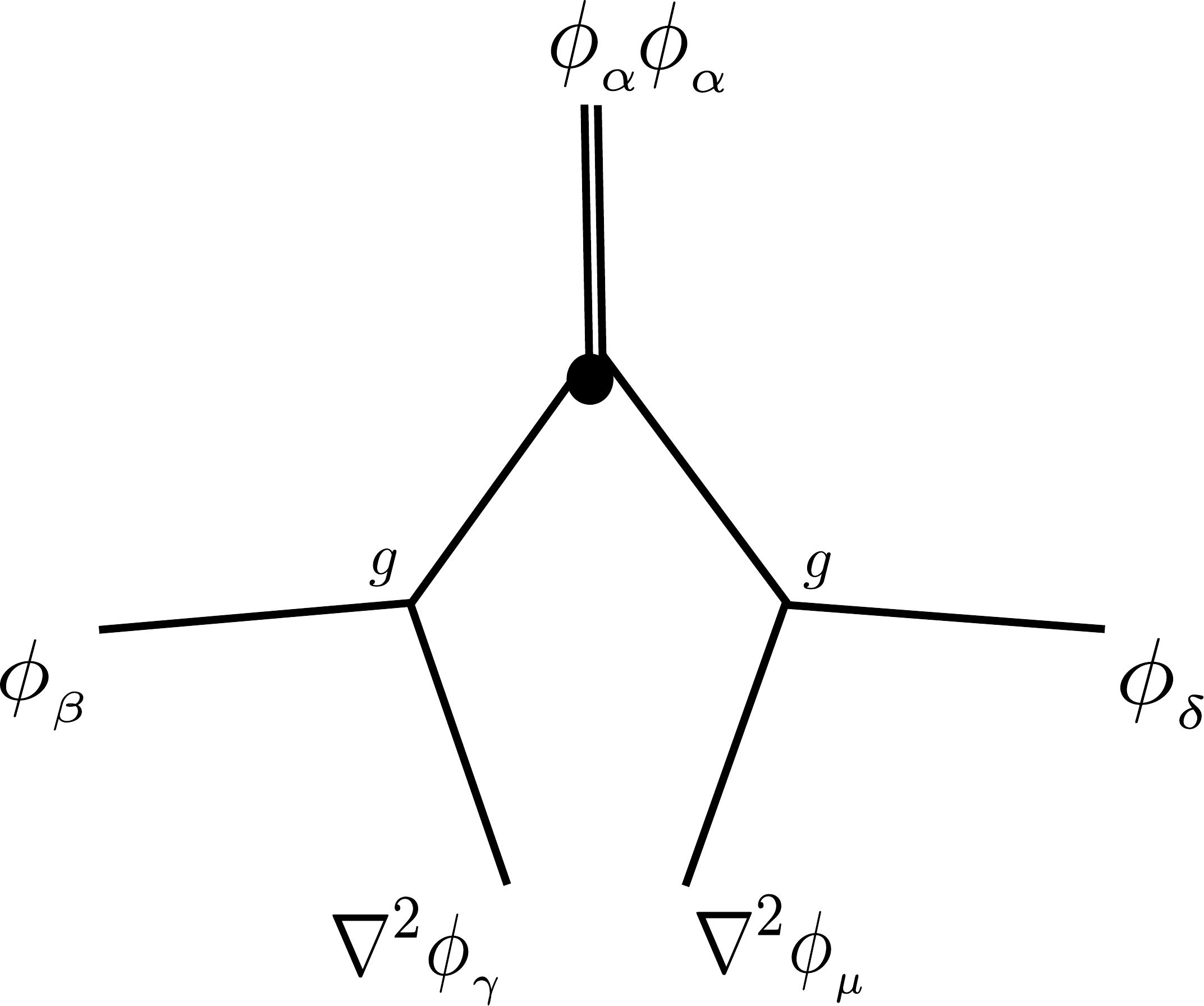}
 \caption{Generation of ${\cal \tilde O}_4({\bf x})$ from ${\cal \tilde O}_2({\bf x})$ in the second order perturbation expansion.}\label{higher1}
\end{figure}
Due to the particular form of the mode-coupling nonlinear term, it is however clear that the fourth order composite operators are generated as well; this, however, has the form $\phi_\alpha({\bf x})(\nabla^2\phi_\beta({\bf x})) \phi_\gamma({\bf x})\nabla^2\phi_\delta({\bf x})$, which should be subleading to $\phi_\alpha({\bf x})\phi_\beta({\bf x}) \phi_\gamma({\bf x})\phi_\delta({\bf x})$ for any $\alpha,\beta,\gamma,\delta$. We thus expect this issue of the generation of the higher order composite operators is unimportant in the asymptotic long wavevelength limit. Nonetheless, this issue continues to persist in the realistic problems of hydrodynamic turbulence and its analogue systems. We do not discuss it here any further.

\section{Multiscaling in Model II}\label{model-II}

Let us now consider the multiscaling in Model II, where we set $\tilde \Gamma(q)=\eta q^2$. We work in the large $\eta$ limit, in which the velocity dynamics effectively reduces to the Stokes' equation. In this limit, the Galilean invariance is restored. We focus on $\overline y < -2$. We closely follow the calculational scheme outlined for Model I above; see also Ref.~\cite{tirtha}. Model II is Galilean invariant and as a result, $\lambda$ does not renormalise at any order in the perturbation theory. Furthermore, there are no renormalisations to $g,\overline D$ and $D_\eta$ as well, for reasons identical to those in Model I.
It is clearly only the diagrams that originate from the advective vertex in (\ref{action1}) are going to be affected by the choice $\tilde \Gamma(q)=\eta q^2$. In other words, only the flow equation for $\nu$ is affected. Noting that the infrared divergence those Feynman graphs, for the same choice of $y$, are now more than what they were with $\tilde\Gamma(q)=\Gamma$. In fact, it is easy to see read these contributions from the corresponding contributions in Model I by replacing $y$ by $y+2$. This, after performing simple and standard procedure, gives
\begin{equation}
  \frac{d\tilde g_2}{dl}=\tilde g_2\left[y+2-d -\frac{g_1}{d}- 2\tilde g_2\frac{d-1}{d}\right],
\end{equation}
where $\tilde g_2\equiv k_d\frac{\lambda^2 D_\eta}{\nu^2}\Lambda^{d-y-2}$. The flow equation for the other dimensionless coupling constant $g_1$ is still given by (\ref{flow-g1}) with $g_2$ replaced by $\tilde g_2$. The fixed points can be directly read off the fixed points obtained for Model I: (i) (0,0), (ii) $(\frac{2d}{3}(\overline y + 2 -d),\, 0)$, (iii) $(0,\,\frac{d}{2(d-1)}(y+2-d))$, and (iv) $2\frac{g^*_1}{4}+\tilde g^*_2(d-1)=d(y+2-d)=2(\overline y-d+2)\frac{d}{3}$. 

We focus on the fixed point $g_1^*>0,\,\tilde g_2^*=0$. We note that the condition of the  nonrenormalisation of $\overline D$ (or equivalently the flow equation (\ref{flowD})) still holds giving $\chi=\delta/3$, as in Model I together with $2\delta/3 > \tilde\epsilon \equiv y+2-d$. At this fixed point, the mode coupling term is relevant and, as in Model I, the structure functions ${\cal \tilde S}_{2n}(r)$ characterise the universality in the inertial range. Unsurprisingly, ${\cal \tilde S}_{2n}(r)$ show only scaling identical to Model I. At the other fixed point $g_1^*=0,\tilde g_2^*>0$, the mode coupling term is irrelevant (in a RG sense) and the structure functions ${\cal S}_{2n}(r)$ are the ones which describe the universal multiscaling with exponents identical to those in Ref.~\cite{tirtha}.


\begin{thebibliography}{99}
\bibitem{nu-comm} In the equilibrium limit of the model, $\nu \propto T-T_c>0$, where $T_c$ is the critical temperature. Thus, we are considering the ``paramagnetic phase''.
\bibitem{u-term} We have ignored a cubic term of the form $\phi_\beta\phi_\beta\phi_\alpha$ for simplicity.
\end{thebibliography}
\end{document}